\documentclass[pre,twocolumn,superscriptaddress,showpacs]{revtex4}
\usepackage{amssymb,amsmath,yfonts}
\usepackage{bm}
\usepackage{psfrag}
\usepackage{xspace}
\usepackage{ulem} 
\usepackage{ae,aecompl} 
\usepackage[T1]{fontenc}
\usepackage[latin1]{inputenc}

\newlength{\maxfigwidth}
\usepackage[dvips]{graphicx}
\usepackage[hypertex]{hyperref}
\usepackage{calc}
\newlength{\dsubscrlena}\newlength{\dsubscrlenb}
\newcommand{\dsubscr}[2]{
  \settowidth{\dsubscrlena}{\widthof{$\scriptstyle{#1}$}}
  \settowidth{\dsubscrlenb}{\widthof{$\scriptstyle{#2}$}}
  \ifdim\dsubscrlena<\dsubscrlenb
  \setlength{\dsubscrlena}{\dsubscrlenb}
  \fi
  \begin{minipage}[b]{\dsubscrlena}
    $\scriptstyle{#1}$ \\[-0.8ex] $\scriptstyle{#2}$ 
  \end{minipage}
}

\newcommand{\eg}{{\itshape e.g.\/}\xspace}
\newcommand{\ie}{{\itshape i.e.\/}\xspace}
\newcommand{\etal}{{\itshape et al.\/}\xspace}

\newcommand{\I}{\textup{i}}
\newcommand{\D}{\textup{d}}

\newcommand{\lqq}{``}
\newcommand{\rqq}{''\xspace}

\newcommand{\ee}{\mathrm{e}} \newcommand{\Mat}[1]{\mathbf{#1}\,}
 \renewcommand{\vec}[1]{\bm{#1}}

\newcommand{\osum}[1]{\operatorname*{\overline{\sum}}_{#1}}
\newcommand{\psum}[1]{\operatorname*{\sum^\prime\nolimits\!}_{#1}\,}
\newcommand{\olsum}[1]{\operatorname*{\overline{\sum}}_{#1}\limits}
\newcommand{\plsum}[1]{\operatorname*{\sum^\prime\nolimits\!}_{#1}\limits\,}

\newcommand{\Sgl}{S_\textup{gl}}
\newcommand{\Ssc}{S_\textup{sc}}
\newcommand{\Svr}{S_\textup{var}}

\newcommand{\lambdaprep}{\lambda_\textup{p}}
\newcommand{\lambdameas}{\lambda_\textup{m}}
\newcommand{\lambdaeff}{\lambda_\textup{eff}}
\newcommand{\chiprep}{\chi_\textup{p}}
\newcommand{\chimeas}{\chi_\textup{m}}
\newcommand{\chicrit}{\chi_{\textrm{crit}}}
\newcommand{\chitrans}{\chi_\textup{t}}
\newcommand{\lenprep}{\xi_{\textrm{p}}}
\newcommand{\lenmeas}{\xi_{\textrm{m}}}
\newcommand{\lenloc}{\xi_{\textrm{l}}}
\newcommand{\scalf}{w}
\newcommand{\kcrit}{k_\textup{c}}
\newcommand{\kmin}{k_\textup{min}}

\begin{document}
\title{Glassy correlations and microstructures in randomly crosslinked
  homopolymer blends} %
\date{\today} %
\author{Christian Wald} %
\affiliation{Institut f\"ur Theoretische Physik, Georg-August-Universit\"at
  G\"ottingen, Friedrich-Hund-Platz 1, 37077 G\"ottingen} %
\email{wald@theorie.physik.uni-goettingen.de}
\author{Paul M. Goldbart} %
\affiliation{Department of Physics, University of Illinois at
  Urbana-Champaign, 1110 West Green Street, Urbana, IL 61806-3080} %
\author{Annette Zippelius} %
\affiliation{Institut f\"ur Theoretische Physik, Georg-August-Universit\"at
  G\"ottingen, Friedrich-Hund-Platz 1, 37077 G\"ottingen} %

\begin{abstract}
  
  We consider a microscopic model of a polymer blend that is prone to phase
  separation. Permanent crosslinks are introduced between randomly chosen
  pairs of monomers, drawn from the Deam-Edwards distribution. Thereby, not
  only density but also concentration fluctuations of the melt are quenched-in
  in the gel state, which emerges upon sufficient crosslinking. We derive a
  Landau expansion in terms of the order parameters for gelation and phase
  separation, and analyze it on the mean-field level, including Gaussian
  fluctuations. The mixed gel is characterized by thermal as well as
  time-persistent (glassy) concentration fluctuations. Whereas the former are
  independent of the preparation state, the latter reflect the concentration
  fluctuations at the instant of crosslinking, provided the mesh size is
  smaller than the correlation length of phase separation. The mixed gel
  becomes unstable to microphase separation upon lowering the temperature in
  the gel phase. Whereas the length scale of microphase separation is given by
  the mesh size, at least close to the transition, the emergent microstructure
  depends on the composition and compressibility of the melt. Hexagonal
  structures, as well as lamell\ae\ or random structures with a unique
  wavelength, can be energetically favorable.

\end{abstract}

% \pacs{82.70.Gg}{Gels and sols}
% \pacs{64.75.+g}{Solubility, segregation, and mixing; phase separation}
% \pacs{61.43.-j}{Disordered solids}

\maketitle

\section{Introduction}

Crosslinked homopolymer blends exhibit a rich phase diagram, due to the
competition between phase separation and crosslinking. The simplest case is of
a blend of two homopolymer species, \lqq{\itshape A\/}\rqq and \lqq{\itshape
  B\/}\rqq, whose incompatibility is controlled by the Flory-Huggins parameter
$\chi$ and which are crosslinked irreversibly by some number $N_c$ of chemical
bonds. In addition, the concentration fluctuations can be controlled
independently in the process of crosslinking (preparation state) and the well
crosslinked gel (measurement state), \eg by lowering the temperature in the
gel.  Hence we have three control parameters: the incompatibility $\chiprep$
in the preparation state, the incompatibility $\chimeas$ in the measurement
state, and the number of crosslinks per chain, $\mu=N_c/N$, where $N$ denotes
the total number of chains in the melt.

A statistical mechanical theory thus has to include not only the average over
the quenched disorder (crosslink realization) but also the \lqq memory\rqq of
the preparation state. This can be achieved in the following way: We start
from a microscopic model, which accounts for the repulsive interaction of all
monomers, irrespective of species (excluded volume), as well as for a
repulsive interaction between the different species only (incompatibility).
Crosslinks are introduced between randomly-chosen pairs of monomers. The
probability for a particular crosslink configuration depends on the
\emph{preparation state} of the system, such that monomers with a high
probability to be close in the preparation state have a high probability to be
crosslinked. Thereby the crosslinks indeed preserve the memory of the
preparation state. Mathematically, this is achieved via the Deam-Edwards
distribution~\cite{deam76} and the replica trick to average over the quenched
disorder.

We expect and indeed find signatures of the preparation state in the gel. An
example are the concentration fluctuations which are frozen in by the
crosslinks. If the preparation state is close to macroscopic phase separation
then the glassy, \ie time persistent, concentration fluctuations reflect the
correlations of the melt at the moment of crosslinking. If, on the other hand,
the preparation state is far from phase separation then the frozen
fluctuations are completely random and follow the pattern set by the
crosslinks.

Lowering the temperature in the gel or, equivalently, increasing the
incompatibility at measurement $\chimeas$ will ultimately give rise to
microphase separation, while macroscopic demixing is suppressed by the
crosslinks. A variety of microphases can exist, depending on the composition
of the blend and its compressibility. If the mixture is symmetric, having
equal concentrations of \lqq{\itshape A\/}\rqq and \lqq{\itshape B\/}\rqq
monomers, then lamellae are energetically favorable, whereas for an
asymmetric mixture hexagons prevail. A finite compressibility enhances the
tendency towards phase separation and can induce a random pattern, consisting
of a superposition of many lamellar phases of different orientation. In all
cases the critical wave-number is given by the mesh size or localization length
of the gel.

The subject of crosslinked homopolymer blends was first addressed by de
Gennes~\cite{genn79}, who pointed out that the mixed state is stabilized in
the gel and eventually undergoes microphase separation. His predictions were
verified experimentally~\cite{brib88,jinn93} with, however, a discrepancy in
the scattering intensity for small wave-number. This was traced back to the
neglect of concentration fluctuations which are present during crosslinking
and are partially frozen in by the crosslinks.  Subsequently several attempts
were made to include these effects
approximately~\cite{brib88,bett91,benm94,read95}.Studies of crosslinked
systems, based on the microscopic model by Panyukov and Rabin~\cite{pany96a},
were reported by Sfatos and Shakhnovich~\cite{sfat97}; as far as homopolymer
blends are concerned, these authors recover de Gennes' result within a
microscopic approach.  Computer simulations were carried out by Lay and
Sommer~\cite{lay_00} who studied, in particular, the relation of the domain
sizes to the mesh size of the gel.  In the last section of our paper we
present a detailed discussion of the literature in comparison with our own
results.

The paper is organized as follows: In sec.~\ref{sec:model} we formulate a
microscopic model of crosslinked polymer chains. Subsequently
(sec.~\ref{sec:free-energy}) we derive a Landau expansion in terms of the
order parameters for gelation and phase separation. The Landau theory allows
us to discuss the mixed gel (sec.~\ref{sec:mixed-state}) as well as microphase
separation (sec.~\ref{sec:microstructures}).  We conclude with a short
summary, a comparison with previous theoretical work and an outlook. A short
account of our results for the special case of an incompressible melt with
equal concentrations of {\itshape A\/}\ and {\itshape B\/}\ monomers was given
previously in ref.~\cite{wald05}.

\section{Model}
\label{sec:model}

\subsection{Uncrosslinked homopolymer blend}

We first consider an \emph{uncrosslinked} blend of polymer, modeled as a
system of Gaussian phantom chains of equal degree of polymerization~$L$ and
step length~$b$.  The melt is taken to contain~$N_A$ chains of type A
and~$N_B$ chains of type B. In general, there will be an imbalance in
concentration $q:=(N_A-N_B)/N$ where $N=N_A+N_B$ is the total number of
chains, occupying a volume~$\mathcal V$ in $d$-dimensional space.  The
monomers positions are denoted as $\vec{R}_{a,i}(s)$, where $a=A,B$ refers to
the chain species, $i=1,\ldots,N_a$ enumerates the chains, and $s=0\dots 1$ is
the continuous index for a \lqq site\rqq on a chain. It turns out to be
convenient to express the monomer positions by dimensionless vectors
$\vec{r}_{a,i}(s) = \sqrt{d\,/\,Lb^2} \; \vec{R}_{a,i}(s)$, so that all
lengths are measured in units of the radius of gyration of the free chains,
$R_{\textrm{g}}=Lb^2/6$.  The rescaled volume reads
$V:=(d/Lb^2)^{d/2}\;\mathcal{V}$.

The chain connectivity is described by the usual Wiener Hamiltonian
\begin{equation}
  H^{W} = \frac{k_{\textrm{B}} T}{2} \sum_{a=A,B} \sum_{i=1}^{N_a}
  \int_0^1\!\D s
  \left(\frac{\D\vec{r}_{a,i}(s)}{\D s}\right)^2\,,
\end{equation}
the excluded volume term controlling compressibility reads
\begin{multline}
  H^\lambda = \frac{V k_{\textrm{B}} T\lambda}{4N} \sum_{a,a^\prime=A,B}
  \sum_{i,i^\prime=1}^{N_a} \int_0^1\!\D s\int_0^1\!\D s^\prime
  \\ \times
  \delta\big(\vec{r}_{a,i}(s)-\vec{r}_{a^\prime,i^\prime}(s^\prime)\big)\,,
\end{multline}
and the incompatibility of the two monomer species is modeled by the
interaction
\begin{multline}
  H^\chi = -\frac{V \chi}{4N}
  \sum_{a,a^\prime=A,B} (2\delta_{a,a^\prime}-1) \sum_{i,i^\prime=1}^{N_a}
  \int_0^1\!\D s\int_0^1\!\D s^\prime
  \\ \times
  \delta\big(\vec{r}_{a,i}(s)-\vec{r}_{a^\prime,i^\prime}(s^\prime)\big)\,.
\end{multline}
Although the chain elasticity and the volume exclusion are of mainly entropic
origin, the incompatibility is assumed to be a chiefly energetic contribution.
Nevertheless we let $k_\textup{B} T=1$ in the following to simplify the
expressions. Instead of changing the temperature, we shall tune $\lambda$ and
particularly $\chi$, which will serve as the inverse temperature.

\subsection{Crosslinking}
Chemical crosslinking induces a random number $M$ of permanent bonds between
randomly selected pairs of monomers; a particular realization of crosslinks is
denoted by ${\cal C}=\{(a_e,i_e,s_e,$\hspace{0ex}$a^\prime_e, i^\prime_e,
s^\prime_e)\}_{e=1}^M$. The links are modeled as hard constraints with zero
bond length. The partition function of the crosslinked melt, relative to a
melt of non-interacting chains, thus reads
\begin{widetext} \begin{align}
\label{eq:1}
  Z({\cal C})
  &:= \biggl\langle \,\prod_{e=1}^M \delta\left( \vec{r}_{a_e,i_e}(s_e) 
    -\vec{r}_{a^\prime_e,i^\prime_e}(s^\prime_e)
  \right) \exp\left\{-H^\lambda-H^\chi\right\} \biggr\rangle^{W} \nonumber \\
  &:= \frac{\int\mathcal D\vec{r}_{a,i}(s)\ 
    \prod_{e=1}^M \delta\left( \vec{r}_{a_e,i_e}(s_e) 
      -\vec{r}_{a^\prime_e,i^\prime_e}(s^\prime_e) \right)
    \ \exp\left\{-H^W-H^\lambda-H^\chi\right\}}
  {\int\mathcal D\vec{r}_{a,i}(s)\,\exp\left\{-H^W\right\}}\,.
\end{align} \end{widetext}
Here we have implicitly defined the expectation value $\langle\dots\rangle^W$
with respect to the Hamilton function of the uncrosslinked melt.

\subsection{Disorder average and Deam-Edwards distribution}
We are interested in the properties of the {\lqq}generic{\rqq} melt rather
than in the properties of a melt with a specific set of crosslinks.
Furthermore, we assume the system to be self-averaging in the thermodynamic
limit. Therefore we will consider \emph{disorder averages} of the observables
with respect to the quenched randomness of crosslinks.

We specify the probability distribution of the crosslink sets following the
strategy of Deam and Edwards~\cite{deam76}. We suppose that the dominant
crosslink sets are those that are most compatible with the uncrosslinked
melt. More precisely, we assume a probability distribution
\begin{equation} \label{eq:2}
  \mathbb{P}_M
  ({\cal C})  \propto  \frac{(\mu N/V)^M}{M!}
  \, Z_\textup{p} ({\cal C})\,.
\end{equation}
Here, $Z_\textup{p}$ is given by eq.~(\ref{eq:1}), evaluated at
$\lambda=\lambdaprep$ and $\chi=\chiprep$, which characterize the system prior
to crosslinking.  Disorder averages with respect to $\mathbb{P}_M$ will be
denoted by square brackets.

\subsection{Order parameters for the homopolymer blend}
\label{sec:order-param-homop} 

To discriminate between the liquid state and the amorphous solid state of the 
polymer system we use the order parameter proposed in~\cite{gold96}:
\begin{multline} 
\label{eq:3}
  \tilde{\Omega}_{\vec k_1\;\dots\;\vec k_g} := \frac{1}{N}
  \sum_{a=A,B}\sum_{i=1}^{N_a}\int_0^1\!\D s\,
  \big\langle \exp\big(\I\vec k_1\vec{r}_{a,i}(s)\big)\big\rangle_{\cal C} \\
  \times 
  \dots \times
  \big\langle \exp\big(\I\vec k_g\vec{r}_{a,i}(s)\big)\big\rangle_{\cal C}\,,
\end{multline}
for $g=1,2,\dots$ and nonzero $\{\vec k_\gamma\}$. The symbol $\langle
\,\dots\, \rangle_{\cal C}$ denotes the thermal expectation value in the
presence of a particular realization ${\cal C}$ of crosslinks. In the case
$g=1$, eq.~(\ref{eq:3}) is the thermal average of the monomer density in
Fourier space,
\begin{equation} \label{eq:4}
  \tilde{\rho}_{\vec k}  :=  \frac{1}{N}\sum_{a=A,B} \sum_{i=1}^{N_a}
  \int_0^1\!\D s\,
%\big\langle
  \ee^{-\I\vec k\vec{r}_{a,i}(s)}
%\big\rangle_{\cal C}
  \,.
\end{equation}

In the liquid state, a monomer explores the sample volume uniformly. Hence,
the equilibrium value of the local density is constant and the Fourier
transform $\langle \exp(i\vec k\vec{r}_i(s))\rangle_{\cal C}$ vanishes (except
for $\vec k=\vec 0$, which we exclude). The order parameter~(\ref{eq:3})
therefore is always zero in the liquid state.

In a solid, at least a finite fraction of the monomers are localized about
points $\vec{b}_{a,i}(s)$ in space. For these monomers, $\langle\exp(\I\vec
k\vec{r}_{a,i}(s))\rangle_{\cal C} \propto
\ee^{\I\vec{k}\vec{b}_{a,i}(s)}\neq0$. However for an amorphous, \ie
macroscopically translationally invariant (MTI) solid, the disorder averaged
expectation value $[\tilde{\Omega}_{\vec k_1\;\dots\;\vec k_g}]$ vanishes
unless $\vec k_1+ \dots+ \vec k_g=0$, see~\cite{gold96}. In particular
$[\langle\tilde{\rho}_{\vec k}\rangle_{\cal C} ]=0$ in the MTI state.  Hence
we can discriminate between the liquid and the amorphous solid state by means
of the $g\geq2$ values of eq.~(\ref{eq:3}).
(For the signature of crystalline and globular states, see~\cite{gold96}).

Throughout this article, we will also refer to the two monomer species as \lqq
opposite charges\rqq. The identification of {\itshape A\/}s and {\itshape
  B\/}s with positive and negative charges, respectively, leads to a natural
choice for an order parameter detecting phase separation: the \lqq charge
density\rqq
\begin{multline}\label{eq:5}
  \tilde{\Psi}_{\vec k} := \frac{1}{N}\sum_{i=1}^{N_A} \int_0^1\!\!\D s\,
  \ee^{-\I\vec k\vec{r}_{A,i}(s)} -\frac{1}{N}\sum_{i=1}^{N_B} \int_0^1\!\!\D
  s\, \ee^{-\I\vec k\vec{r}_{B,i}(s)}
\end{multline}
measuring the local imbalance of the concentrations of $A$ and $B$.

In the general case of an asymmetric blend, in which there is an excess of
either {\itshape A\/}- or {\itshape B\/}-chains, the average charge density is
given by $q\cdot N/V=(N_A-N_B)/V$, so that $\tilde{\Psi}_{\vec k}$ serves as
an order parameter only for ${\vec k}\neq 0$. Homogeneous phase separation is
indicated by a nonzero expectation value of the order parameter in the limit
$k \to 0$.  Microstructures, \eg lamellae or hexagonally ordered cylinders,
give rise to a nonzero expectation value of $[\langle\tilde{\Psi}_{\vec
  k}\rangle_{\cal C}] $ at finite wave-number. %
A nonuniform charge density in general is accompanied by mass density
modulations, except for the incompressible case.

In the gel phase we expect to find static charge fluctuations
$\langle\tilde{\Psi}_{\vec k}\rangle_{\cal C}\neq 0$ for all wave-numbers. If
the gel state is a homogeneous mixture of $A$ and $B$ chains then the disorder
averaged charge density vanishes, \ie $[\langle\tilde{\Psi}_{\vec
  k}\rangle_{\cal C}]= 0$.  The frozen-in fluctuations can only be detected by
the glassy correlations $[\langle\tilde{\Psi}_{\vec k}\rangle_{\cal
  C}\langle\tilde{\Psi}_{-\vec k}\rangle_{\cal C}]$.  In general, the
quadratic expectation values $[\langle \tilde{\Psi}_{\vec k}
\tilde{\Psi}_{-\vec k} \rangle_{\cal C}]$ and $[ \langle \tilde{\Psi}_{\vec k}
\rangle_{\cal C} \langle \tilde{\Psi}_{-\vec k} \rangle_{\cal C} ]$ measure
volatile and time-persistent charge fluctuations in an a priori homogeneous
mixture.

\section{Effective free energy}
\label{sec:free-energy}

The disorder averaged free energy 
\begin{equation}
  F =-\big[\ln Z({\cal C})\big] %
  = -\lim_{n\to0} \frac{\big[\bigl(Z\bigr)^n\big] - 1}{n} 
\end{equation}
is computed with help of the replica trick. %
The $n$-th power of $Z$ is made explicit using $n$ independent copies --
\emph{replicas}~-- of the system and an additional replica is introduced to
account for the Deam-Edwards distribution~\cite{gold96}. The average over the
disorder can be carried out explicitly, yet at the cost of a coupling between
the formerly independent replicas, yielding
\begin{equation}
  \label{eq:7}
  \big[Z^n\big] =: \mathcal{Z}_n/\mathcal{Z}_0 \,,
\end{equation}
where the replicated partition function is given by
\begin{widetext} \begin{equation}
  \label{eq:8} %\textstyle 
  \mathcal{Z}_n
  = \left\langle \exp\Bigg\{ -H^\lambda_{n+1}-H^\chi_{n+1}
    +\frac{\mu V}{4N}
    \!\!\sum_{a,a^\prime=A,B}\limits\, \sum_{i,i^\prime=1}^{N}\limits
    \int_0^1\limits \!\D s\,\D s^\prime \prod_{\alpha=0}^n\limits
    \delta\bigl( \vec{r}_{a,i}^{\,\alpha}(s)
    -\vec{r}_{a^\prime,i^\prime}^{\,\alpha}(s^\prime)
    \bigr)\Bigg\} \right\rangle^W_{n+1} \ .
\end{equation} \end{widetext}
Here, the $\vec r^{\alpha}_{a,i}(s)$ denote the monomer positions in the
$\alpha^{th}$ replica, $\langle\, \dots\, \rangle^W_{n+1}$ is the replicated
Wiener average, and $H^\lambda_{n+1}$ and $H^\chi_{n+1}$ denote the replicated
Hamiltonians of the excluded volume and incompatibility interactions. The
denominator and the zeroth replica in the numerator are due to the
Deam-Edwards distribution~(\ref{eq:2}) and reflect the situation \emph{prior}
to crosslinking.  Thus, we have to distinguish between the zeroth replica,
characterized by $\lambdaprep$ and $\chiprep$ \emph{(preparation ensemble)},
and the other $n$ replicas, reflecting the situation \emph{after
  crosslinking}, characterized by $\lambdameas$ and $\chimeas$
\emph{(measurement ensemble)}. To account for the particular role of the
zeroth replica we use the notation
\begin{align}
  \lambda^\alpha & := \begin{cases} \ \lambdaprep & \textrm{if }
    \alpha=0,
    \\[1ex]
    \ \lambdameas & \textrm{otherwise,}
  \end{cases} &&\text{and}\\
  \chi^{\alpha} & := \begin{cases}
    \ \chiprep & \textrm{if } \alpha=0,
    \\[1ex]
    \ \chimeas & \textrm{otherwise.}
  \end{cases}
\end{align}

The many-particle problem of the polymer melt can be formally reduced to a
two-chain problem (one chain of each species).  It is convenient to introduce
$(n+1)$-fold replicated vectors $\hat{x} := \big( \vec{x}^0, \dots,
\vec{x}^n\big)$, and to express the exponents in eq.~(\ref{eq:8}) in Fourier
space, which leads to terms quadratic in the monomer and charge densities.
These can be linearized by means of Hubbard-Stratonovich transformations,
yielding
\begin{equation} \label{eq:9}
  \mathcal{Z}_n = \mathcal{B}_n \,\cdot\,
  \int\!\mathcal{D}(\{\Psi\,,\Omega\,,\rho \})
  \exp\bigl\{-n\mathcal{F}_n(\{\Psi,\Omega,\rho\}) \bigr\}
\end{equation}
with a constant $\mathcal{B}_n= \exp\big\{ \tfrac{N}{2}
  (-n\lambdameas+nq^2\chimeas+((n+1)V^{-n}-1)\mu) \big\}=1+\mathcal{O}(n)$
and the \emph{effective free energy}
\begin{widetext}\begin{multline} \label{eq:10} % \textstyle
    n \mathcal{F}_n(\{\Psi,\Omega\,\rho\}) = \frac{N}{2} \plsum{\alpha,\vec
      k}\left(\! \left(\frac{1}{\chi^\alpha}-
        \frac{q^2}{\tilde{\lambda}^\alpha}\right) \bigl|\Psi^\alpha_{\vec
        k}\bigr|^2 + \frac{1}{\tilde{\lambda}^\alpha} \bigl|\rho^\alpha_{\vec
        k}\bigr|^2 +
      \frac{\I\,q}{\tilde{\lambda}^\alpha}\big(\rho^\alpha_{\vec
        k}\Psi^\alpha_{-\vec k}
      +\rho^\alpha_{-\vec k}\Psi^\alpha_{\vec k}\big)\!\right) \\
    +\frac{NV^{n}}{2\mu} \olsum{\hat k}\limits \bigl|\Omega_{\hat k}\bigr|^2
    \textstyle -N_A \ln z_{+}\big(\{\Psi,\Omega,\rho\}\big) -N_B \ln
    z_{-}\big(\{\Psi,\Omega,\rho\}\big)\,,
\end{multline} 
where
\begin{equation} %multline} 
  z_{\pm}(\{\Psi,\Omega,\rho\}) := % \textstyle \\
  \biggl\langle \exp\biggl\{ \plsum{\alpha,\vec k} \big( +\I
    \rho^\alpha_{\vec k}-(q \mp 1) \Psi^\alpha_{\vec k} \big) %\\ \times
  \int_0^1\!\D s\, \ee^{\I\vec k\vec r^\alpha(s)} + V^{-n} \olsum{\hat k}
  \Omega_{\hat k} \cdot \int_0^1\!\D s\, \ee^{\I\hat{k}\hat{r}(s)}
  \biggr\}\biggr\rangle^W_{n+1}\,.
\end{equation} %multline}
\end{widetext}
The Wiener average now runs over a \emph{single replicated chain} having
monomer positions $\hat{r}(s)$. The auxiliary fields $\Psi_{\vec{k}}$,
$\rho_{\vec{k}}$ and $\Omega_{\hat{k}}$ are pairwise dependent via
$\Omega_{-\hat{k}} = (\Omega_{-\hat{k}})^*$ etc., thus the $\Psi$ and $\Omega$
integrations in~(\ref{eq:9}) are restricted to the half spaces $\vec{k} \cdot
\vec{n}>0$ and $\hat{k} \cdot \hat{n}>0$ (with arbitrary nonzero constants
$\vec{n}$ and $\hat{n}$). The sums over the $\vec{k}^\alpha$ and $\hat{k}$ are
split into {\em replica sectors\/}: The constant part $\mathcal{B}_n$ is
composed of the $\vec{k}^\alpha =\vec{0}$ and $\hat{k}=\hat{0}$ contributions
({\em zero replica sector\/}). The symbols $\psum{\vec{k}}$ and
$\osum{\hat{k}}$ denote the sums over nonzero $\vec{k}$ ({\em single replica
  sector\/}) and over $\hat{k}$ with nonzero $\vec{k}^\alpha$ in at least two
replicas ({\em higher replica sector\/}), respectively; both restricted to the
above subspaces. 

As expected, crosslinks give rise to an attractive interaction between the
chains, which, in the absence of excluded volume, would cause the chains to
collapse into a globular state. A sufficiently strong excluded volume
interaction prevents this collapse, as can be read off the coefficient
$\tilde{\lambda}^{\alpha}:=\lambda^{\alpha}-\mu/V^n$ of the
term quadratic in the density. Stability requires $\lambdaprep>\mu$ and
$\lambdameas>\mu$.

The physical observables, moments of the local density and charge
density, are related to expectation values of the fields according to
\begin{align*}
  \bigl[\bigl\langle \tilde{\Psi}_{\vec k} \bigr\rangle_{\cal
    C}\bigr]&\,=\,\tfrac{1}{\chimeas} \lim_{n\rightarrow0}\limits
  \bigl\langle\Psi^\alpha_{\vec k}\bigr\rangle^{\mathcal{F}}_{n+1}
  &&\text{and}\\
  \bigl[\bigl\langle \tilde{\rho}_{\vec k} \bigr\rangle_{\cal
    C}\bigr]&\,=\,\tfrac{\I}{\lambdameas-\mu} \lim_{n\rightarrow0}\limits
  \bigl\langle\rho^\alpha_{\vec k} +\I q
  \Psi^\alpha_{\vec k} \bigr\rangle^{\mathcal{F}}_{n+1}  \\
  \intertext{for $\vec{k}\neq\vec{0}$ and $\alpha\geq1$, and} \bigl[
  \tilde{\Omega}_{\vec k_1\;\dots\;\vec k_g} \bigr]&\,=\,\tfrac{1}{\mu}
  \lim_{n\rightarrow0}\limits \bigl\langle\Omega_{\hat
    k}\bigr\rangle^{\mathcal{F}}_{n+1}
\end{align*}
for $\hat{k}=\bigl(\vec{0}, \vec{k_1}, \dots, \vec{k}_g, \vec{0}, \dots,
\vec{0}\bigr)$ with $g\geq2$.

In asymmetric blends, which have an excess of either {\itshape A\/} or
{\itshape B\/} chains, the average charge density $qN/V=(N_A-N_B)/V$ is
nonzero. To simplify the Landau expansion of the free energy it is then
advantageous to either work with the fluctuations $\delta \Psi =\Psi-q$ of the
charge density around its mean value or, alternatively, shift the monomer
density, as is done here.

\section{Homogeneously mixed states}
\label{sec:mixed-state}

On the mean field level, we approximate the functional integral over $\Omega$,
$\rho$ and $\Psi$ in~(\ref{eq:9}) by using the saddle point method, \ie by the
value of the integrand at the point $(\bar{\Psi},\bar{\rho},\bar{\Omega})$
making the integrand stationary:
\begin{equation}
  \mathcal{Z}_n \ \sim\  \textup{const}\,\cdot\,
 \ee^{- \mathcal{F}_n\big(\bar{\Psi},\bar{\rho},\bar{\Omega}\bigr)}
\end{equation}
where, by definition, $(\bar\Psi,\bar{\rho},\bar\Omega)$ satisfy the
stationarity conditions
\begin{align}
  \label{eq:18}
  \frac{\partial \mathcal{F}_n}{\partial \Psi^\alpha_{\vec k}}
  \biggr|_{\bar\Psi,\bar\rho,\bar\Omega} & = 0 \,, &
  \frac{\partial \mathcal{F}_n}{\partial \rho^\alpha_{\vec k}}
  \biggr|_{\bar\Psi,\bar\rho,\bar\Omega} & = 0 \,,
  %&&\text{and}&
  &\frac{\partial \mathcal{F}_n}{\partial \Omega_{\hat k}}
  \biggr|_{\bar\Psi,\bar\rho,\bar\Omega} & = 0 \,.
\end{align}

\subsection{Homogeneously mixed liquid state} \label{sec:homog-mixed-liqu}
One solution of the stationarity conditions~(\ref{eq:18}) is the trivial
saddle point $\bar{\Psi}=\bar{\rho}=\bar{\Omega}=0$, corresponding to the
homogeneously mixed liquid state. To assess its stability we consider the
Landau expansion to leading order around this point:
\begin{widetext} \begin{multline} \label{eq:19}
  \tfrac{2n}{N} \mathcal{F}_n(\{\Psi,\rho,\Omega\}) = \plsum{\alpha,\vec k}
  \Bigl( \tfrac{1}{\tilde{\lambda^\alpha}} + g_\textup{D}\bigl(\vec
  k^2\bigr)\!\Bigr) \big| \rho^\alpha_{\vec k} \big|^2 + \olsum{\hat k} \Bigl(
  \tfrac{1}{\mu} - g_\textup{D}\bigl(\hat k^2 \bigr)\!\Bigr)
  \big| \Omega_{\hat k} \bigr|^2  \\
  + 2\I q\plsum{\alpha,\vec{k}} \tfrac{1}{\tilde{\lambda}^\alpha}\,
  \Omega^\alpha_{-\vec k}\Psi^\alpha_{\vec k} + \plsum{\alpha,\vec{k}} \Bigl(
  \tfrac{1}{\chi^\alpha} - \tfrac{q^2}{\tilde{\lambda}^\alpha} -(1-q^2)
  g_\textup{D}\bigl(\vec k^2\bigr)\!\Bigr) \big| \Psi^\alpha_{\vec k} \big|^2
  +{\cal O}(\Psi^2,\rho^2,\Omega^2)\,,
\end{multline} \end{widetext}
where the Debye function~$g_\textup{D}$ is defined in
appendix~\ref{sec:corr-funct}. The stability limits of the homogeneous liquid
can be read off from the quadratic coefficients. As~$g_\textup{D}(k^2)$
decreases monotonically from one to zero, stability against solidification and
demixing require $\mu<1$ and $(\chiprep,\chimeas)<1/(1-q^2)$, respectively.
Throughout this article, we assume that $(\lambdaprep,\lambdameas)>\mu$, \ie
that the excluded volume interaction is strong enough to prevent density
instabilities (see above).

It should be noted that the condition $(\chiprep,\chimeas)<1/(1-q^2)$ denote
local stability \emph{limits} only. In mean-field theory, the phase separation
transition for symmetric blends is of second order, so the phase transition
coincides with the limit of local stability. In asymmetric blends the
transition is of first order. The loss of local stability, as given by the
conditions above, then defines a spinodal, and the transition occurs at a
lower value of~$\chi$. The location of the spinodal depends on the average
charge~$q$, with a larger critical incompatibility (\ie lower critical
temperature) for more asymmetric mixtures.

The gelation of the homogeneous liquid, driven by increasing the crosslink
concentration, and the microphase separation of the resulting gel, induced by
cooling, will be addressed in the following sections.

\subsection{Crosslinking in the homogeneously mixed state
  \label{sec:crossl-homog-mixed}} 
In the liquid state, the polymer blend phase separates macroscopically beyond
the demixing threshold. The subsequent gelation of such a macrophase-separated
melt would result, apart from the interface, in just two pieces of gel having
different compositions.  It is more interesting to consider a gel prepared
from a homogeneous melt to study phase separation in the gel phase.  As we
shall see below, such a gel shows glassy charge density patterns and, as
anticipated, microphase separation instead of macroscopic demixing. Therefore,
the discussion will be restricted to crosslinking in a homogeneously mixed
blend, $\chiprep<1/(1-q^2)$, including undercooled mixtures for $q\neq0$.

Upon gelation, the saddle point $\bar\Omega_{\hat{k}}=0$ will become unstable,
making it necessary to complement the expansion~(\ref{eq:19}) of the free
energy with the third-order terms:
\begin{widetext} \begin{multline} \label{eq:20}
    \tfrac{2n}{N} \mathcal{F}_n(\{\Psi,\rho,\Omega\}) = \plsum{\alpha,\vec k}
    \Bigl( \tfrac{1}{\chi^\alpha} -(1-q^2) g_\textup{D}\bigl(\vec
    k^2\bigr)\!\Bigr) \big| \Psi^\alpha_{\vec k} \big|^2 + \olsum{\hat k}
    \Bigl( \tfrac{1}{\mu} - g_\textup{D}\bigl(\hat k^2 \bigr)\!\Bigr)
    \big| \Omega_{\hat k} \bigr|^2  \\
    -\sum_{\alpha_1\neq\alpha_2} \plsum{\vec k_1,\vec k_2} \olsum{\hat{p}}\,
    \Omega_{\hat{p}} \Psi^{\alpha_1}_{\vec k_1}\, \Psi^{\alpha_2}_{\vec k_2}\,
    \delta_{\vec{p}^{\alpha_1},\vec{k}_1}\delta_{\vec{p}^{\alpha_2},\vec{k}_2}
    - \tfrac{1}{3}\olsum{\hat{k}_{1,2,3}}
%  g_3\left(\hat{k}_1,\hat{k}_2\right)\,
    \Omega_{\hat{k}_1}\Omega_{\hat{k}_2}\Omega_{\hat{k}_3}
    \,\delta_{\hat{k}_1+\hat{k}_2+\hat{k}_3,\hat{0}}\,.
\end{multline} \end{widetext}
The vertex functions of the cubic terms have been approximated by their zero
wave-number values, the complete expressions being given in
appendix~\ref{sec:corr-funct}.  This approximation is well justified, because
the gelation transition is always continuous, so that the relevant
length-scales are very large, compared with the scales of the microscopic
correlations. Here, we have taken the limit of an incompressible melt, which
is achieved by integrating out the density fluctuations on the Gaussian level
and subsequently taking the limit $\tilde{\lambda}\to\infty$.

We first discuss a gel in the homogeneously mixed state $\bar\Psi^\alpha_{\vec
  k}=0$, assuming $\chimeas<1/(1-q^2)$. Following~\cite{gold96}, we consider
the order parameter hypothesis
\begin{equation}
    \label{eq:21} \bar\Omega_{\hat k} = \delta_{\tilde{\vec{k}},\vec 0}\cdot Q
  \int_0^\infty\!\D\tau\,p(\tau)\exp(-\tilde{\vec{k}}^2/2\tau)
\end{equation}
with the shorthand $\tilde{\vec{k}}:=\sum_{\alpha=0}^n \vec{k}^{\alpha}$.
Here, $Q$ denotes the fraction of chains that are localized, \ie the \emph{gel
  fraction}; the localization lengths are distributed according to the
distribution function $p(\tau)$. Both have to be determined self-consistently
as a solution of the stationarity conditions~(\ref{eq:18}).

The first two of the stationarity conditions are satisfied for any $Q$ and
$p(\tau)$. The third condition is independent of the incompatibility
parameter. Hence, in the homogeneously mixed regime, the task of determining
$Q$ and $p(\tau)$ on the saddle-point level is exactly the same as in the pure
gelation problem in~\cite{gold96}. In the present notation, the result for the
solid state, \ie $\mu>1$, reads
\begin{multline}
  \label{eq:22}
  \bar\Omega_{\hat k} \ \approx\ 
  \delta_{\tilde{\vec{k}},\vec 0} \cdot 2\mu(\mu-1) \cdot \omega\left(\sqrt{
      \tfrac{4}{3}\, \tfrac{\hat{k}^2}{\mu-1}} \;\right) \\
  =\,\delta_{\tilde{\vec{k}},\vec 0} \cdot \mu(\mu-1) \cdot
  w\left( \hat{k}^2/2(\mu-1)\right)\,,
\end{multline}
with the gel fraction approximately given by $Q\approx 2(\mu-1)$. In the
liquid state, $\bar\Omega_{\hat{k}}=0$ and $Q=0$.  The scaling function
$\omega(x)$ is defined in~\cite{gold96} (see appendix~\ref{sec:interp-form}
for details); for convenience we define the shorthand $w(x) := \omega \left(
  \sqrt{8x/3} \right)$.

\subsection{Stability of the homogeneously mixed gel}
\label{sec:stability}
Starting from a gel prepared from a homogeneous melt, \ie
$\chiprep<1/(1-q^2)$, we now allow the incompatibility to be changed after
crosslinking.  In order to keep the gel homogeneous, $\chimeas$ must remain
smaller than the critical value $\chicrit$ for (micro-)phase separation. As can
be seen from the term coupling $\Psi_{\vec{k}}$ and $\Omega_{\hat{k}}$ in the
effective free energy~(\ref{eq:20}), the gel network stabilizes the mixed
state; the details are discussed in the following.

As long as the gel is homogeneous, the order parameter~(\ref{eq:22}) solves
the stationarity conditions~(\ref{eq:18}). To determine the stability of the
mixed state, we need the second derivative of~$\mathcal{F}_n$ with respect to
the charge density, evaluated at the saddle point.  We restrict the discussion
to a weak gel, \ie $\mu-1\ll1$, so that the saddle point value of
$\Omega_{\hat k}$ is small and the Hessian can be approximated by its
expansion to linear order in~$\bar\Omega_{\hat k}$.  It then can be read off
from the Landau expansion (\ref{eq:20}) with $\Omega_{\hat{k}}$ replaced by
the explicit saddle point value~(\ref{eq:22}).  The latter vanishes unless
$\tilde{\vec k}=0$: hence there is no coupling between the
different~$\vec{k}$'s, and the Hessian can be calculated independently for
each wave-vector. We obtain
\begin{equation} \label{eq:24}
  \frac{\partial^2\mathcal{F}_n}
  {\partial\Psi^{\alpha_1}_{\vec k} \partial\Psi^{\alpha_2}_{-\vec k}}
  \biggr|_{\bar\Psi,\bar\Omega}
  \approx\ N(1-q^2) \cdot A_{\alpha_1\,\alpha_2}(k) \,,
\end{equation}
where
\begin{equation} \label{eq:25}
  A_{\alpha_1\,\alpha_2} := \begin{pmatrix}
    c & -b & \cdots & -b \\
    -b & a & \ddots & \vdots \\
    \vdots & \ddots & \ddots & -b \\
    -b & \hdots & -b & a 
  \end{pmatrix}
\end{equation}
with
\begin{align} \label{eq:6}
  a &:= \bigl(\tfrac{1}{(1-q^2)\chimeas}
  -g_\textup{D}\bigl(k^2 \bigr) \bigr)\,, \nonumber\\[1ex]
  b &:= \mu(\mu-1) \,
  w\bigl(k^2/(\mu-1)\bigr) \,, &\text{and} \nonumber\\[1ex]
  c &:= \bigl( \tfrac{1}{(1-q^2)\chiprep}
  -g_\textup{D}\bigl(k^2 \bigr) \bigr)\,.
\end{align}
  
The stability of the homogeneous state is equivalent to the positivity of
$\Mat{A}$. In the limit $n\rightarrow0$, its eigenvalues are given by
\begin{align} \label{eq:38}
  \lambda_1(k) &:=c && \text{(non-degenerate),} &\text{and} \\
  \label{eq:42} \lambda_2(k) &:=a+b && \text{($n$-fold degenerate)}.
\end{align}
As we assume crosslinking in the mixed phase, $\lambda_1$ is always positive,
and thus the stability condition reduces to $\lambda_2>0$ or, equivalently
$\chimeas<\chicrit(\mu)$ with
\begin{multline} 
  \label{eq:26}
  (1-q^2)\chicrit(\mu) := \\ 
  1/ \max_{k} \left\{ \, g_\textup{D}\big( k^2
    \big) -\mu(\mu-1) \,
%  g_{\Psi^2\Omega}\big( k^2 \big) 
  w\big( k^2/(\mu-1) \big) \right\} \,.
\end{multline}

\begin{figure} \center
  \psfrag{PFmu0.00}[Bl]{\large$\!\mu\leq1$}
  \psfrag{PFmu1.01}[Bl]{\large$\!\mu=1.01$}
  \psfrag{PFmu1.02}[Bl]{\large$\!\mu=1.02$}
  \psfrag{PFmu1.05}[Bl]{\large$\!\mu=1.05$}
  \psfrag{PFmu1.10}[Bl]{\large$\!\mu=1.10$}
  \psfrag{PFkqu}[cl]{\large$k^2$}
  \psfrag{PFlb2}[tc][][1][0]{\mbox{}\large\raisebox{10mm}{$\lambda_2(k)$}}
  \psfrag{PY0.00}[Bl]{\normalsize\ \ $0.00$} \psfrag{PY0.05}[Bl]{\normalsize\ 
    \ $0.05$} \psfrag{PY0.10}[Bl]{\normalsize\ \ $0.10$}
  \psfrag{PY0.15}[Bl]{\normalsize\ \ $0.15$} \psfrag{PY0.20}[Bl]{\normalsize\ 
    \ $0.20$} \psfrag{PX0.0}[Bl]{\normalsize\ $0.0$}
  \psfrag{PX0.2}[Bl]{\normalsize\ $0.2$} \psfrag{PX0.4}[Bl]{\normalsize\ 
    $0.4$} \psfrag{PX0.6}[Bl]{\normalsize\ $0.6$}
  \psfrag{PX0.8}[Bl]{\normalsize\ $0.8$}
  
  \includegraphics[width=\maxfigwidth]{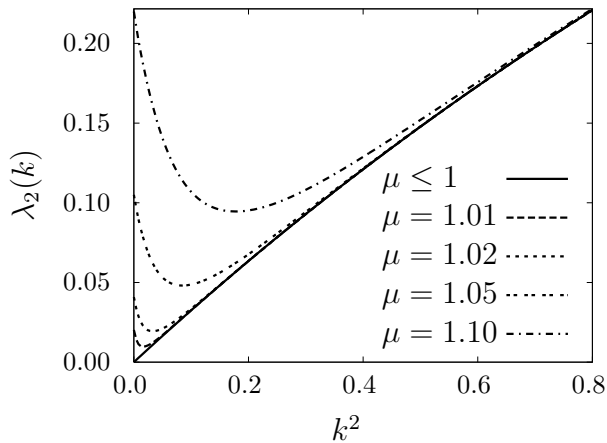}
  \caption{ \label{fig:Stability_a-b}
    Stability parameter $\lambda_2(k)$ for $\chimeas=1/(1-q^2)$ as a
    function of $k^2$.  }
\end{figure}

Figure~\ref{fig:Stability_a-b} shows $\lambda_2(k)$ for $\chimeas=1/(1-q^2)$
and different crosslink concentrations.  Increasing $\chimeas$ shifts the
curve downwards.  Apparently, an instability towards demixing first occurs for
a nonzero wave-number $\kcrit$, which maximizes the above expression.

We consider the case of a weak gel, so $\mu(\mu-1) \approx \mu-1$, and we can
write $g_\textup{D}(k^2) \approx 1-k^2/3$ as the Debye function decays much
more slowly than $w\big( k^2/ (\mu-1) \big)$. In this approximation, $\kcrit$
is given by
\begin{equation}
  0 = \left. \frac{\partial \lambda_2}{\partial k^2}\right|_{k=\kcrit}
  \approx 
    \frac{1}{3}+w^\prime\bigl( \kcrit^2/(\mu-1)\bigr)\,,
\end{equation}
which leads to $\kcrit^2 \approx 1.61(\mu-1)$ and $(1-q^2)\chicrit-1 \approx
\kcrit^2/3+(\mu-1)w\bigl(\kcrit^2/(\mu-1)\bigr) \approx 0.98(\mu-1)$.

A more precise numerical analysis without these approximations yields 
\begin{multline}
  \label{eq:28}
  (1-q^2)\chicrit-1 = \\ 
  0.98\cdot(\mu-1)+0.70\cdot(\mu-1)^2
  +\mathcal{O}\left((\mu-1)^3\right)
\end{multline}
and
\begin{equation}
  \label{eq:29}
  \kcrit^2 = 1.61\cdot(\mu-1) +1.75\cdot(\mu-1)^2
  +\mathcal{O}\left((\mu-1)^3\right)\,.
\end{equation}

The instability for nonzero $k$ implies that the gel undergoes microscopic
(rather than macroscopic) phase separation. This is to be expected, because
crosslinks permanently connect different chains and thus prevent true
macroscopic phase separation. The \lqq next best\rqq state for the system is
phase separation up to the length-scale of the network, \ie the typical mesh
size, as given by the average localization length $\bar\xi\sim(\mu-1)$.  Hence
the instability occurs at a critical wave-number $\kcrit\sim 1/\bar\xi$.

\begin{figure}
  \psfrag{PX-1}[Bl]{\normalsize\ \,-1} \psfrag{PX0}[Bl]{\normalsize\ \,0}
  \psfrag{PX1}[Bl]{\normalsize\ \,1} \psfrag{PY0}[Bl]{\normalsize\ \ \ 0}
  \psfrag{PY1}[Bl]{\normalsize\ \ \ 1} \psfrag{PY2}[Bl]{\normalsize\ \ \ 2}
  \psfrag{PY3}[Bl]{\normalsize\ \ \ 3} \psfrag{PY4}[Bl]{\normalsize\ \ \ 4}
  \psfrag{Pq}[Bl]{\large$q$}
  \psfrag{Pc}[tl]{\large$\chicrit$}
  \psfrag{Pm1}[Bl]{\large\hspace{-4.5ex}$\mu<1$}
  \psfrag{Pm2}[Bl]{\large\hspace{-4.5ex}$\mu=1.1$}
  \psfrag{Pm3}[Bl]{\large\hspace{-4.5ex}$\mu=1.5$}
  \includegraphics[width=\maxfigwidth]{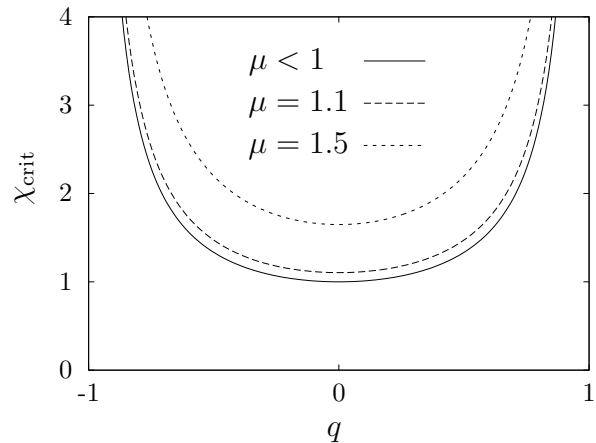}
  
  \caption{ \label{fig:hpb-spinodals-asymm}
    Stability limits $\chicrit(\mu)$ as a function of the asymmetry~$q$ for
    the liquid ($\mu<1$) and gels with different strengths ($\mu=1.1$ and
    $\mu=1.5$).}
\end{figure}

The instability is hampered by an increased density~$\mu$ of crosslinks and
the asymmetry~$q$ of the composition. The spinodals for the liquid blend and
two solid gels with two different degrees of crosslinking are shown in
fig.~\ref{fig:hpb-spinodals-asymm} as a function of $q$. In contrast, the
critical wave-number remains unchanged in agreement with the above argument~--
the critical wave-number is determined by the mesh size, which is unaffected by
$q$. The microphase transition is addressed in section~\ref{sec:micr-separ},
where it will be shown that the average charge also influences the observed
microstructure.

\subsection{Pseudo phase diagram}
Three parameters determine the state of the system: $\mu$ controls the number
of crosslinks, $\chiprep$ specifies the charge fluctuations at preparation and
$\chimeas$ the charge fluctuations after the gel has been prepared. Each of
them can be chosen such that the system is close to a critical point: $\mu=1$
corresponds to the gelation transition, $\chiprep=1/(1-q^2)$ to macroscopic
phase separation in the preparation ensemble and $\chimeas=\chicrit$ to
microphase separation in the gel.

\begin{figure} \center
  \psfrag{YY}{\large\;$\mu$} \psfrag{Y1}{\normalsize\ \,1}
  \psfrag{X0}{\normalsize\ \,0} \psfrag{X1}{\normalsize 1}
  \psfrag{XX}{\large$\chimeas$} \center
  \includegraphics[width=0.8\maxfigwidth]{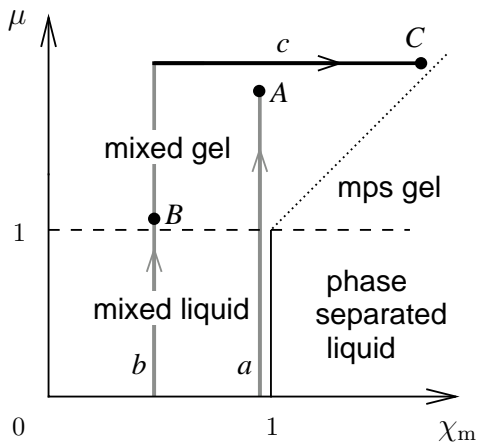}
  
  \caption{ \label{fig:Phase-diagram} 
    Pseudo phase diagram of the polymer blend in the $\chi$-$\mu$-plane. The
    state of the system is, however, history-dependent (see text for
    details).}
\end{figure}

In fig.~\ref{fig:Phase-diagram} we show a phase diagram in the
$\chimeas$-$\mu$ plane for the special case $q=0$. [The spinodals of the
asymmetric case can be recovered by replacing $\chi$ by $(1-q^2)\chi$.] The
dashed line $\mu=1$ separates the gel state and the liquid state. The latter
is further divided into a mixed and a macroscopically phase separated liquid
at $\chimeas=1$ (solid line).  The dotted line $\chimeas=\chicrit$ separates
the mixed gel from the microphase separated one. 

The diagram in fig.~\ref{fig:Phase-diagram} is not a true equilibrium phase
diagram, because the state of the system also depends on the preparation
ensemble via $\chiprep$. In particular, the microphases are only obtained if
crosslinking takes place in the homogeneously mixed phase. As a consequence,
the transition line $\mu=1$ cannot be crossed from the macro- to the
microphase separated state for $\chimeas>1$.
The history of the gel is indicated by a \emph{path} in the diagram.  Of
particular interest are the three paths \emph{a}, \emph{b} and \emph{c}.  Path
\emph{a} amounts to crosslinking close to macroscopic phase separation,
$\chiprep\to 1$, and the endpoint \emph{A} corresponds to a homogeneously
mixed gel having long-ranged frozen-in charge fluctuations.  Along
path~\emph{b} the system is crosslinked in a preparation state that is far
away from macroscopic phase separation. The endpoint \emph{B} corresponds to a
homogeneously mixed and rather weak gel, just crosslinked enough to be
solid-like. Along path~\emph{c}, the system is prepared in the same way as on
path \emph{b}, however more crosslinks are introduced, which strengthens the
gel.  Subsequently, the temperature is lowered (the incompatibility~$\chimeas$
increased), so that the endpoint \emph{C} is close to microphase separation.
These three histories are representative in the following sense. Each endpoint
corresponds to a state close to one critical point, as discussed above: point
\emph{A} is close to macroscopic phase separation in the preparation ensemble,
$\chiprep=1$; point \emph{B} is close to the gelation transition, $\mu=1$; and
point \emph{C} is close to microphase separation, $\chimeas=\chicrit$.  The
three states of the system, corresponding to the endpoints, will be discussed
in detail in the following sections.

\subsection{Charge density correlations in the mixed gel}
\label{sec:charge-dens-corr}
In this section we discuss the homogeneously mixed gel phase, for which the
order parameter for phase separation vanishes. Nevertheless, there are thermal
as well as quenched charge fluctuations on various length-scales. These can be
detected with help of multiple correlation functions.
On the Gaussian level of approximation, these correlation functions are given
by the inverse of the Hessian matrix~(\ref{eq:25}):
\begin{equation}\label{eq:30}
  \bigl\langle \Psi^{\alpha_1}_{-\vec k} \Psi^{\alpha_2}_{\vec k}
  \bigr\rangle^{\mathcal{F}}_{n+1}
  = (A^{-1})_{\alpha_1\,\alpha_2}
  \,.
\end{equation}
The correlator that is off-diagonal in replica space accounts for the
frozen-in correlations and will be termed the \emph{glassy} correlator. It is
given by
\begin{multline}
 \label{eq:33}
 \Sgl(k):=\big[\bigl\langle \tilde{\Psi}_{-\vec k}\bigr\rangle \bigl\langle
 \tilde{\Psi}_{\vec k} \bigr\rangle\big]= \lim_{n\rightarrow 0}\, \bigl\langle
 \Psi^{\alpha}_{\vec k} \Psi^{\beta}_{-\vec k} \bigr\rangle \\
 =\tfrac{1}{\chimeas^2} \cdot \tfrac{b(b+c)}{c\cdot\lambda_2^2}\,;
\end{multline}
see eq.~\ref{eq:5} for the definition of $\tilde{\Psi}$.  The replica-diagonal
correlator is the \emph{scattering intensity}
\begin{multline} 
  \label{eq:36}
  \Ssc(k):=\big[\bigl\langle \tilde{\Psi}_{-\vec k} \tilde{\Psi}_{\vec k}
  \bigr\rangle\big]= \lim_{n\rightarrow 0}\,
  \bigl\langle \Psi^{\alpha}_{\vec k} \Psi^{\alpha}_{-\vec k} \bigr\rangle \\
  =\tfrac{1}{\chimeas^2} \left( \tfrac{b(b+c)}{c\cdot\lambda_2^2} +
    \tfrac{1}{\lambda_2} - \chimeas \right)\,,
\end{multline}
and the \emph{variance} (or connected correlator) is given by
\begin{equation} \label{eq:37}
  \Svr(k):=\Ssc(k)-\Sgl(k)
  =\tfrac{1}{\chimeas^2}
  \cdot \left( \tfrac{1}{\lambda_2} -\chimeas \right) \,.
\end{equation}
Whereas the \emph{glassy correlator}~$\Sgl(k)$ describes the static, frozen-in
correlations, the \emph{variance} $\Svr(k)$ quantifies the volatile, thermal
fluctuations about the mean value. The \emph{scattering intensity} $\Ssc(k)$
is the sum of both contributions and covers both, thermal and static charge
inhomogeneities.

\subsubsection*{Restriction to symmetric blends}
In the following discussion of $\Ssc(k)$, $\Sgl(k)$ and~$\Svr(k)$ we shall
confine ourselves to the case of symmetric blends, yet without loss of
generality: The scattering functions of the asymmetric case are recovered via
multiplication by $\gamma=(1-q^2)$ and the rescalings
$\chiprep\to\gamma\chiprep$ and $\chimeas\to\gamma\chimeas$. Furthermore, the
distance to phase separation is replaced with the distance to the spinodal in
the asymmetric case. In the range between the equilibrium phase transition and
the spinodal, the results then describe an undercooled mixture.

\subsubsection*{Length-scales} \label{sec:hpb-length-scales}

The correlation functions are characterized by three length-scales which are
determined by the parameters ($\mu,\chiprep,\chimeas$) of preparation and
measurement conditions:

First, there is the typical localization length $\xi$ of the monomers in the
gel fraction, \ie the mean mesh size of the gel. From eq.~(\ref{eq:22}) we can
infer that this length-scale is roughly given by
\begin{equation}
  \label{eq:134}
  \lenloc := 1\,/\,\sqrt{\mu-1}.
\end{equation}

Second, there is the decay length $\lenprep$ of the pre-critical demixing
fluctuations prior to gelation. This approximately reads
\begin{equation}
  \label{eq:135}
  \lenprep := 1\,/\,\sqrt{1-\chiprep}.
\end{equation}

The third length characterizes the pre-critical fluctuations of microphase
separation, and is approximately given by
\begin{equation}
  \label{eq:136}
  \lenmeas := 1\,/\,\sqrt{\chicrit(\mu)-\chimeas}\,.
\end{equation}

The three length-scales measure, or are given by, the inverse distance to the
phase transitions of gelation and demixing in the pre-crosslinking blend, and
microphase separation in the gel; hence they grow large when approaching their
respective transitions.  In the following, we shall essentially discuss three
limiting regimes, in which the correlation functions are determined by one of
the three length-scales
\begin{itemize}
\item $\lenprep\gg\lenloc,\lenmeas$ \quad (point \emph{A}),
\item $\lenloc\gg \lenprep,\lenmeas$ \quad (point \emph{B}), \ and
\item $\lenmeas\gg\lenloc,\lenprep$ \quad (point \emph{C}).
\end{itemize}

\subsubsection*{Glassy correlations}

The glassy correlation function $\Sgl(k)$ describes \emph{time-persistent}
charge inhomogeneities due to crosslinking. If the preparation ensemble is
close to phase separation then instantaneous crosslinking will freeze in these
fluctuations, and $\Sgl$ will be dominated by the pre-crosslinking
fluctuations, giving rise to a high value at zero wave-vector. If, on the
other hand, the preparation ensemble is in a well-mixed state then
crosslinking will introduce \emph{completely random, static} charge
fluctuations, which subsequently can be enhanced by approaching the microphase
separation transition in the gel. In the following, we discuss the three
limiting cases (\ie points~\emph{A}, \emph{B}, \emph{C}) in detail.

\begin{figure}
  \psfrag{PY1.2}[Bl]{\normalsize\ \ $1.2$} \psfrag{PY1.0}[Bl]{\normalsize\ \ 
    $1.0$} \psfrag{PY0.8}[Bl]{\normalsize\ \ $0.8$}
  \psfrag{PY0.6}[Bl]{\normalsize\ \ $0.6$} \psfrag{PY0.4}[Bl]{\normalsize\ \ 
    $0.4$} \psfrag{PY0.2}[Bl]{\normalsize\ \ $0.2$}
  \psfrag{PY0.0}[Bl]{\normalsize\ \ $0.0$} \psfrag{PX0}[cl]{\normalsize\,\ 
    $0$} \psfrag{PX1}[cl]{\normalsize\;\ $1$} \psfrag{PX5}[cl]{\normalsize\,\ 
    $5$} \psfrag{PX10}[cl]{\normalsize\,\ $10$}
  \psfrag{PX15}[cl]{\normalsize\,\ $15$} \psfrag{PFY}{}
  \psfrag{PFX}[cc]{\large $k^2\lenprep^2/3$} \psfrag{PSgl}[Bl]{\large$\Sgl$}
  \psfrag{PVar}[Bl]{\large$\Svr$} \psfrag{PSsc}[Bl]{\large$\Ssc$}
      
  \includegraphics[width=\maxfigwidth]{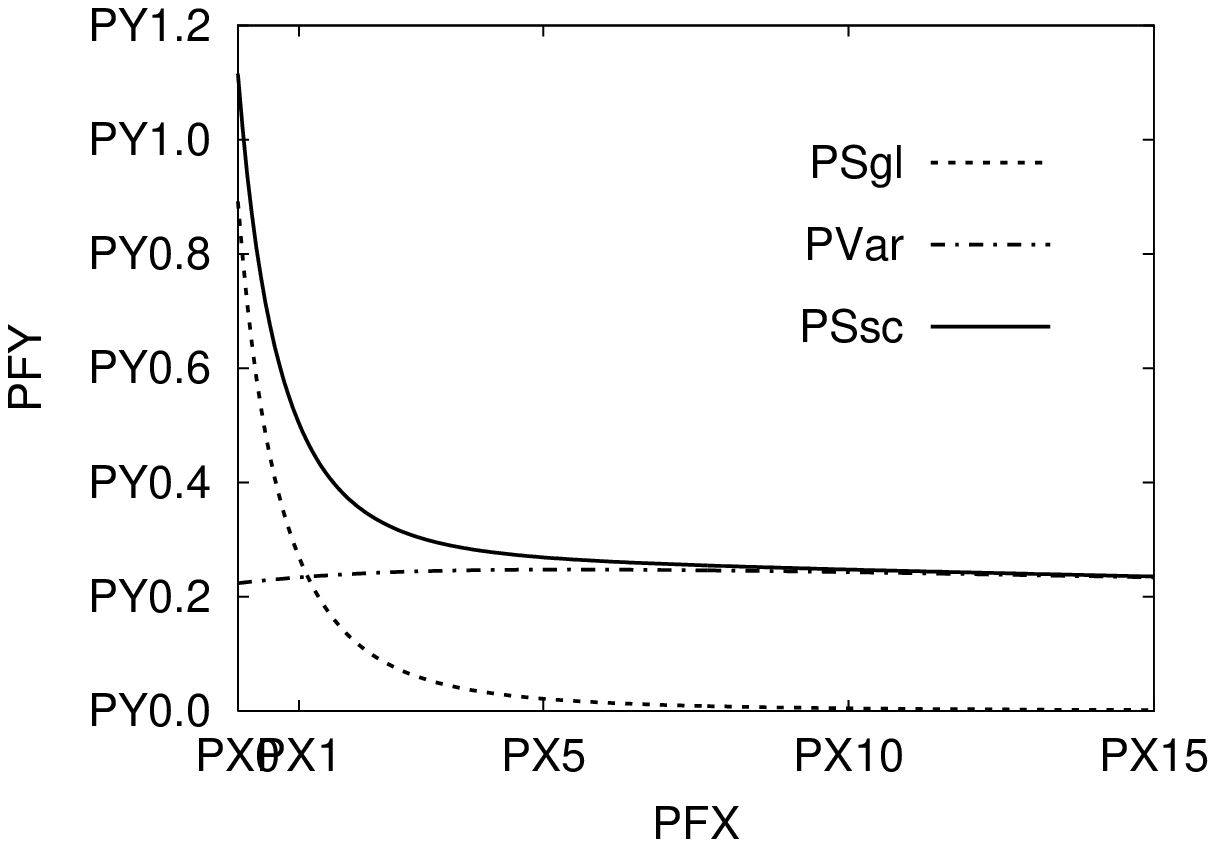}

  \caption{ \label{fig:gel-corr-prepfluct}
    Correlation functions for a ~\emph{gel prepared close to phase
      separation}: $\Sgl$, $\Svr$ and~$\Ssc$ in units of
    $4\lenprep^2\lenmeas^4/\lenloc^4$ for $\lenloc^2=10^2$, $\lenprep^2=10^3$
    and $\lenmeas^2=10$. }
\end{figure}

We first consider a gel that is prepared from a melt close to phase
separation, \ie $\lenprep\gg(\lenloc,\lenmeas)\gg1$, corresponding to point
\emph{A} in fig.~\ref{fig:Phase-diagram}. The network can freeze-in
correlations on length-scales larger than or comparable to its mesh size.  For
$\lenprep\gg \lenloc$, the pre-crosslinking fluctuations have long enough
scales to be frozen. Consequently, the glassy correlations reflect the
pre-crosslinking fluctuations:
\begin{equation}
  \Sgl(k) \propto %\approx 
  \frac{\lenprep^2}{1+k^2\lenprep^2/3 }.
%\cdot \frac{1}
%{\left(c_1^2 + \tfrac{1}{2}\,
%  \lenloc^2/\lenmeas^2\right)^2} 
\end{equation}
%with $c_1^2\approx 0.3$. 
The glassy correlations are proportional to $\lenprep^2$ and decay on the
scale $k\sim\lenprep^{-1}$ set by the fluctuations of the \emph{preparation
  ensemble}. An example is included in fig.~\ref{fig:gel-corr-prepfluct}.

\begin{figure}
  \psfrag{PY1.2}[Bl]{\normalsize\ \ $1.2$} \psfrag{PY1.0}[Bl]{\normalsize\ \ 
    $1.0$} \psfrag{PY0.8}[Bl]{\normalsize\ \ $0.8$}
  \psfrag{PY0.6}[Bl]{\normalsize\ \ $0.6$} \psfrag{PY0.4}[Bl]{\normalsize\ \ 
    $0.4$} \psfrag{PY0.2}[Bl]{\normalsize\ \ $0.2$}
  \psfrag{PY0.0}[Bl]{\normalsize\ \ $0.0$} \psfrag{PX0.0}[cl]{\normalsize\ 
    $0.0$} \psfrag{PX0.5}[cl]{\normalsize\ $0.5$}
  \psfrag{PX1.0}[cl]{\normalsize\ $1.0$} \psfrag{PX1.5}[cl]{\normalsize\ 
    $1.5$} \psfrag{PFY}{} \psfrag{PFX}[cc]{\large $k^2\lenmeas^2/3$}
  \psfrag{PSgl}[Bl]{\large$\Sgl$} \psfrag{PVar}[Bl]{\large$\Svr$}
  \psfrag{PSsc}[Bl]{\large$\Ssc$}
      
  \includegraphics[width=\maxfigwidth]{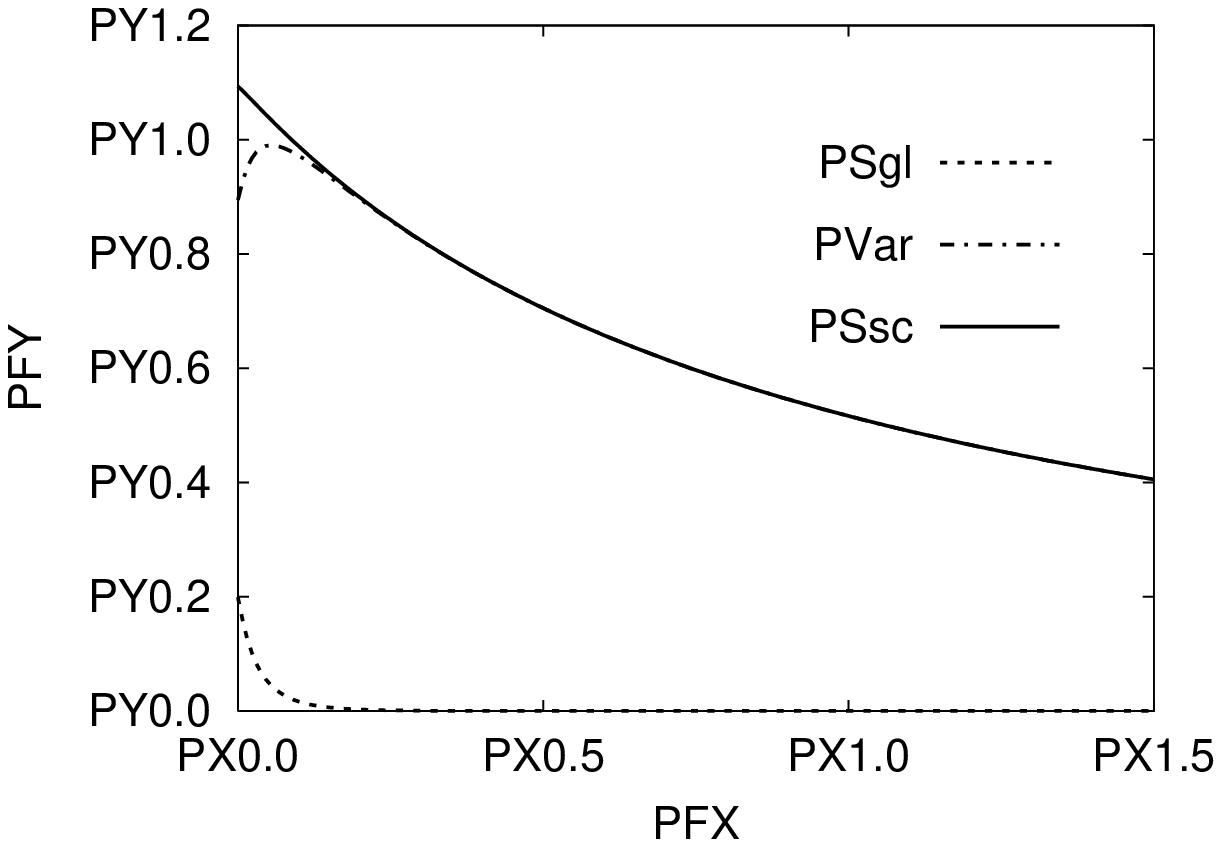}

  \caption{ \label{fig:gel-corr-weakgel}
    Correlation functions for a~\emph{weak gel}: $\Sgl$, $\Svr$ and~$\Ssc$ in
    units of~$\lenmeas^2$ for~$\lenloc^2=10^2$, $\lenprep^2=10$
    and~$\lenmeas^2=10$. }
\end{figure}

In a weak gel, \ie if $\lenloc\gg(\lenprep,\lenmeas)\gg1$ (point \emph{B}),
the network is rather wide-meshed, so that the fluctuations at preparation
cannot be frozen in. Instead, there will be static charge fluctuations
$\langle\tilde{\Psi}_{\vec k}\rangle \neq 0$ on the scale of the network,
which are completely random and hence vanish, if averaged over crosslink
configurations, $[\langle\tilde{\Psi}_{\vec k}\rangle]=0$.  They do, however,
contribute to the glassy correlations, which are given approximately by
\begin{equation}
  \Sgl(k) \approx \tfrac{1}{2}Q\, \lenmeas^4 \cdot
  \scalf(k^2\lenloc^2) \,.
\end{equation}
These fluctuations always decay on the length-scale of localization, but they
are enhanced in magnitude when approaching microphase separation.  An example
of the glassy correlations in this range is given in
fig.~\ref{fig:gel-corr-weakgel}.

\begin{figure}
  \psfrag{Pksq}[Bl]{\large$k^2$}
  \psfrag{PSrel}[bc][bc][0.9]{\raisebox{1ex}{\large$S_{\textrm{gl}}(k)\,/\,S_{\textrm{gl}}(0)$}}
  \psfrag{1e-06}[cl]{\normalsize$10^{-6}$}
  \psfrag{1e-04}[cl]{\normalsize$10^{-4}$}
  \psfrag{1e-02}[cl]{\normalsize$10^{-2}$}
  \psfrag{1e+00}[cl]{\normalsize\,$10^{0}$} \psfrag{0.0}[Bc]{\normalsize\ 
    $0.0$} \psfrag{0.2}[Bc]{\normalsize\ $0.2$} \psfrag{0.4}[Bc]{\normalsize\ 
    $0.4$} \psfrag{0.6}[Bc]{\normalsize\ $0.6$} \psfrag{0.8}[Bc]{\normalsize\ 
    $0.8$} \psfrag{1.0}[Bc]{\normalsize\ $1.0$}
  \psfrag{Plk12}[Bl]{\quad\ $k_{1/2}^2$}
  \psfrag{Pdlc}[bl]{\raisebox{1ex}{$\lenprep^{-2}$}}
  \psfrag{-6P}[Bc][cc]{\footnotesize $\,10^{\mbox{-}6}\,$}
  \psfrag{-4P}[Bc][cc]{\footnotesize $\,\,\,10^{\mbox{-}4}\,$}
  \psfrag{-2P}[Bc][cc]{\footnotesize $\,\,\,\,\,10^{\mbox{-}2}\,$}
  \psfrag{0P}[Bc][cc]{\footnotesize $1$} \psfrag{-4}[cc][cc]{\footnotesize
    $10^{\mbox{-}4}\,$} \psfrag{-2}[cc][cc]{\footnotesize $10^{\mbox{-}2}\,$}
      
  \raisebox{-0.065\maxfigwidth}{
    \includegraphics[height=0.78\maxfigwidth,width=\maxfigwidth]{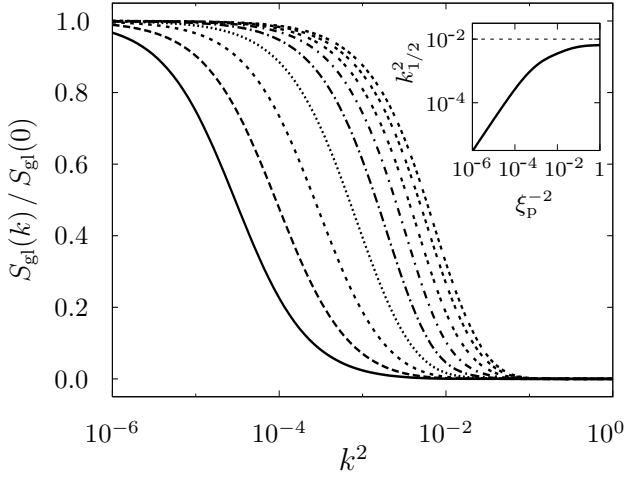}
  }

  \caption{ \label{fig:gel-corr-crossover}
    Scale crossover in the glassy correlation function $\Sgl$, normalized to
    the value at~$k=0$: Crossover from preparation close to demixing to a weak
    gel, with $\chimeas=0.1$, $\lenloc^2=10^2$, and
    $\lenprep^2=10^{5},10^{4.5},\ldots,10^{1}$ (from left to right).  Inset:
    Half-width~$k^2_{1/2}$ of~$\Sgl(k)$, crossing over from $\lenprep^{-2}$ to
    $\lenloc^{-2}$ (dashed line).}
\end{figure}

The cross-over between the two scales is demonstrated in
fig.~\ref{fig:gel-corr-crossover}, which shows $\Sgl(k)/ \Sgl (0)$ far from
microphase separation for $\lenloc=10$.  For the leftmost curve
$\lenprep^2=10^5 \gg \lenloc^2=100$, and hence the decay occurs at $k \sim
\lenprep^{-1}$. Upon decreasing $\lenprep$, the curves shift to the right,
until, for $\lenloc \gg \lenprep$, the decay is determined by $\lenloc$.  The
inset shows the half-width at half-maximum as a function of $\lenprep^{-1}$.

\begin{figure}
  \psfrag{PY0.6}[Bl]{\normalsize\ \ $0.6$} \psfrag{PY0.4}[Bl]{\normalsize\ \ 
    $0.4$} \psfrag{PY0.5}[Bl]{\normalsize\ \ $0.5$}
  \psfrag{PY0.2}[Bl]{\normalsize\ \ $0.2$} \psfrag{PY0.3}[Bl]{\normalsize\ \ 
    $0.3$} \psfrag{PY0.0}[Bl]{\normalsize\ \ $0.0$}
  \psfrag{PY0.1}[Bl]{\normalsize\ \ $0.1$} \psfrag{PX0.0}[cl]{\normalsize\ \ 
    $0.0$} \psfrag{PX0.5}[cl]{\normalsize\ \ $0.5$}
  \psfrag{PX1.0}[cl]{\normalsize\ \ $1.0$} \psfrag{PX1.5}[cl]{\normalsize\ \ 
    $1.5$} \psfrag{PX2.0}[cl]{\normalsize\ \ $2.0$}
  \psfrag{PX2.5}[cl]{\normalsize\ \ $2.5$} \psfrag{PFY}{}
  \psfrag{PFX}[tc]{\large $k^2/\kcrit^2$} \psfrag{PSgl}[Bl]{\large$\Sgl$}
  \psfrag{PVar}[Bl]{\large$\Svr$} \psfrag{PSsc}[Bl]{\large$\Ssc$}
      
  \includegraphics[width=\maxfigwidth]{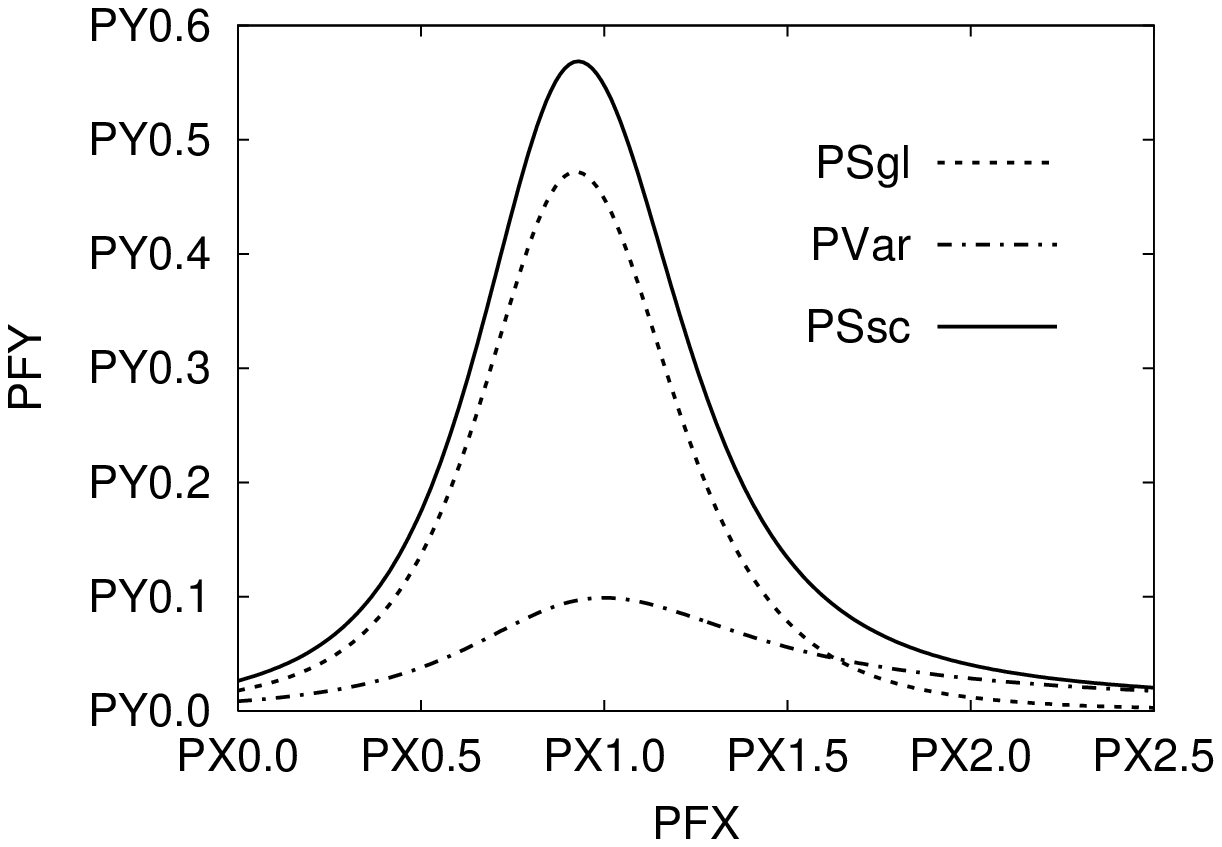}

  \caption{ \label{fig:gel-corr-closemps}
    Correlation functions for a~\emph{gel measured close to microphase
      separation}: $\Sgl$, $\Svr$ and~$\Ssc$ in units
    of~$\lenmeas^4/\lenloc^2$ for~$\lenloc^2=10^2$, $\lenprep^2=10$~and
    $\lenmeas^2=10^3$. The wave-number squares are measured in units
    of~$\kcrit^2$. }
\end{figure}

Close to microphase separation, \ie $\lenmeas\gg (\lenloc, \lenprep) \gg1$
(point \emph{C}), $\Sgl(k)$ is dominated by the critical fluctuations towards
microphase separation.  In the absence of crosslinks there would be
large-scale fluctuations towards macroscopic demixing. In the gel,
displacements are bounded by the localization length, so that $\Sgl(k)$
develops a peak at $\kcrit \sim \lenloc^{-1}$, where $\lambda_2(k)$ becomes
small; an example is included in fig.~\ref{fig:gel-corr-closemps}.
Approaching the transition, the peak diverges as $\lambda_2^{-2}(k)$, and the
glassy correlations can be approximated by
\begin{equation}
  \Sgl(k) \propto \frac{Q}{(\chicrit(\mu)-\chimeas
    +\tfrac{1}{2} (k^2-\kcrit^2)^2 \scalf''(1)/\kcrit^2)^2}\,,
\end{equation}
where $\scalf''$ denotes the second derivative of the scaling function
introduced below eq.~(\ref{eq:22}) and defined in
appendix~\ref{sec:scaling-function}.

\subsubsection*{Thermal fluctuations}

The variance $\Svr(k)$ of the charge fluctuations is \emph{independent} of the
conditions at the time of crosslinking. Hence, there are only two competing
length-scales, $\lenloc$ and $\lenmeas$. In case of a weak gel, \ie
$\lenloc\gg \lenmeas \gg1$ (point \emph{B} in fig.~\ref{fig:Phase-diagram}),
eq.~(\ref{eq:37}) reduces to
\begin{equation}
  \Svr(k) \approx \frac{\lenmeas^2}{1+k^2\lenmeas^2/3}
\end{equation}
for small $k$, decaying with a half-width of $k_{1/2} \approx
\sqrt{3}/\lenmeas^2$, provided $\lenmeas$ is not too small. Far away from the
demixing threshold, the fluctuations are hardly affected by the network and
look like critical fluctuations approaching macroscopic phase separation. An
example is shown in fig.~\ref{fig:gel-corr-weakgel}.

Close to microphase separation, \ie $\lenmeas \gg \lenloc \gg1$ (point
\emph{C}), the fluctuations grow with increasing~$k$ until~$k$ reaches the
inverse localization length $k\sim \lenloc^{-1}$, beyond which they are
strongly suppressed by the network. The variance is approximately given by
\begin{multline}
  \Svr(k)\approx \frac{1}{\lambda_2(k)} \\
  \approx 
  \frac{1}{\chicrit(\mu)-\chimeas
    +\tfrac{1}{2} (k^2-\kcrit^2)^2 \scalf''(1)/\kcrit^2}\,,
\end{multline}
revealing a peak at $k_0\sim1/\lenloc$ that has a height proportional to
$\lenmeas^2$. See fig.~\ref{fig:gel-corr-closemps} for an example.

\subsubsection*{Scattering intensity}
The behavior of the scattering function in the various regimes can be
inferred from the behaviors of~$\Sgl(k)$ and~$\Svr(k)$, as~$\Ssc(k)$ is just
the sum of them. A weak gel (point~\emph{B}) preserves only a small amount of
the pre-crosslinking fluctuations and can hardly restrict thermal
fluctuations.  Hence, $\Ssc(k)\approx \Svr(k)$, so the scattering function
decays on the scale $k\sim \lenmeas$; see fig.~\ref{fig:gel-corr-weakgel}.
In the other regimes, thermal fluctuations are suppressed by the network, and
$\Ssc(k)\approx \Sgl(k)$. In a gel \emph{prepared} close to phase separation
(point \emph{A}), the scattering function decays on the scale $k\sim \lenprep$
(fig.~\ref{fig:gel-corr-prepfluct}), whereas a gel \emph{measured} close to
microphase separation (path \emph{C}) reveals a peak at $k\sim k_0$, diverging
at the transition (fig.~\ref{fig:gel-corr-closemps}).

\section{Microstructures} \label{sec:micr-separ}
\label{sec:microstructures}

At $\chimeas=\chicrit$, the homogeneous gel becomes unstable with respect to
phase separation. As we have seen in section~\ref{sec:stability}, the
instability first occurs for nonzero wave-numbers, indicating that the gel
undergoes microscopic rather than macroscopic phase separation. In this
section we investigate various microstructures, such as hexagons and
lamell\ae\ with a definite orientation, as well as a superposition of many
random orientations.  The selection of a particular microstructure depends
sensitively on the compressibility and the charge imbalance. We first discuss
the simplest case of incompressible, symmetric mixtures, and then go on to
investigate the effects of charge imbalance and compressibility.

\subsection*{Incompressible, symmetric mixtures}
\label{sec:mps-symmetric-case}

Our analysis of microphase separation is based on the effective free
energy~(\ref{eq:10}) in the gel phase. We expand it around $\Psi=0$ up to
fourth order, in the presence of a nonzero gel order parameter $\bar\Omega$
given by eq.~(\ref{eq:22}).  The expansion reads
\begin{widetext} \begin{multline} \label{eq:34}
  \tfrac{n}{N} \mathcal{F}_n\left(\left\{\Psi\right\}\right) \approx
  \tfrac{1}{2} \sum_{\alpha_{1,2}} \plsum{\vec{k}} A_{\alpha_1\,\alpha_2}(k)\,
  \Psi_{\vec{k}}^{\alpha_1}\Psi_{-\vec{k}}^{\alpha_2}
  \\
  +\tfrac{1}{8} \sum_{\alpha} \plsum{\vec k_{1,2,3,4}}\!
  \delta_{\vec{k}_1+\vec{k}_2+\vec{k}_3+\vec{k}_4,\vec{0}} \left(
    \frac{g_3(\vec{k}_1,\vec{k}_2) g_3(\vec{k}_3,\vec{k}_4)}
    {g_\textup{D}((\vec{k}_1+\vec{k}_2)^2)} - \frac{
      g_{\Psi^4}(\vec{k}_1,\vec{k}_2,\vec{k}_3)}{3} \right)
  \Psi^{\alpha}_{\vec k_1}\, \Psi^{\alpha}_{\vec k_2}\, \Psi^{\alpha}_{\vec
    k_3}\, \Psi^{\alpha}_{\vec{k}_4}
\end{multline}
Here, $A_{\alpha_1\,\alpha_2}(k)$ is given in eqs.~(\ref{eq:25},\ref{eq:6}),
and the vertex of the fourth-order term is given by the Wiener correlators
\begin{equation} 
  \label{eq:psi3}
  g_3\big(\vec k_1, \vec k_2\big) =
  \int_0^1\!\D s_1\D s_2\D s_3\,
  \left. \Bigl\langle
    \ee^{-\I\sum_{\nu=1}^3 \vec k_\nu\vec r(s_\nu)}
    \Bigr\rangle^W \right|_{\vec k_3=-\vec k_1-\vec{k}_2}
\end{equation} 
and 
\begin{equation} 
\label{eq:psi4}
  g_{\Psi^4}\big(\vec k_1, \vec k_2, \vec k_3 \big) =
  \int_0^1\!\D s_1\D s_2\D s_3\D s_4\,
  \left. \Bigl\langle
    \ee^{-\I\sum_{\nu=1}^4 \vec k_\nu\vec r(s_\nu)}
    \Bigr\rangle^W \right|_{\vec k_4=-\sum_{\nu=1}^3 \vec k_\nu}.
\end{equation}  \end{widetext}

The fourth-order term in eq.~(\ref{eq:34}) apparently depends on the spatial
structure of the microphases, and is responsible for the pattern selection as
well as for the wave-number selection in the microphase separated state.  A
particularly simple pattern are lamell\ae\ with sinusoidal modulations in real
space:
\begin{equation}
  \label{eq:lamella}
  \bar\Psi^{\alpha}_{\vec {p}} = \begin{cases}
    0, & \textrm{for }\alpha=0, \\ 
   \sqrt{2} \bigl( \delta_{\vec{p},\vec k}
    + \delta_{\vec{p}, -\vec k} \bigr)\, \psi,
    & \textrm{otherwise}. \end{cases}
\end{equation}
This ansatz is replica-symmetric, apart from the zeroth replica, which
reflects the preparation ensemble. The wave-length $2\pi/k$ and the
amplitude~$\psi\geq0$ are variational parameters subject to optimization.
Insertion into~(\ref{eq:34}) yields
\begin{multline}
  \label{eq:35}
  f\left(\{\Psi\}\right) := \tfrac{1}{N} \cdot \lim_{n\rightarrow0}
  \mathcal{F}_n(\{\Psi\}) \\
  \approx \lambda_2(k^2)\cdot\psi^2 + \tfrac{1}{2} g_4(k^2) \cdot \psi^4\,,
\end{multline}
with $\lambda_2$ defined in eq.~(\ref{eq:42}) and 
\begin{multline}
  \label{eq:32}
  g_4(k^2) := \frac{1}{2} \left(
    \frac{(g_3(\vec{k},\vec{k}))^2}{g_\textup{D}(k^2)}
    +2g_\textup{D}(k^2) - g_{\Psi^4}(\vec{k},\vec{k},-\vec{k})\!\right) \\
  = 1-\frac{1}{3}k^2 +\mathcal{O} \big(k^4\big)
\end{multline}
(see appendix~\ref{sec:hpb-vertex-funct}).

At the onset of the microphases, the amplitude goes to zero continuously, and
the optimal wave-number is given by $\kcrit$ [see eq.~(\ref{eq:29})], implying
a domain-size of the order of the localization length of the gel.  Beyond the
critical point, the amplitude is nonzero, and the wavenumber deviates from its
critical value~$\kcrit$.
Both are obtained from a variational optimization of the above free energy:
\begin{align}
  \psi^2_{\text{min}}(k)
  &= -\frac{\lambda_2(k)}{g_4(k^2)} \\
  \intertext{and} \kmin^2-\kcrit^2 &= c_0\cdot (\chimeas-\chicrit) \,+\,
  \mathcal{O}\big((\chimeas-\chicrit)^2\big)\,.
\end{align}
The constant $c_0$ can be computed analytically only for a weak gel, by
further expansion in powers of $\mu-1$, which yields~$c_0 \approx 2.03
(\mu-1)$. The optimal amplitude $\psi_{\text{min}}(\kmin)$ grows continuously
with $\chimeas-\chicrit$:
\begin{equation}
  \psi_{\text{min}}^2 = \frac{\chimeas-\chicrit}{\chicrit^2\;g_4(\kcrit^2)}
  \quad +\ \mathcal{O}\left( (\chimeas-\chicrit)^2 \right)\,.
\end{equation}

Other simple structures, such as a hexagonal stack of cylinders or a bcc
crystal of spheres, have higher free energies. The same holds for a
superposition of several sinusoidal modulations like eq.~(\ref{eq:lamella}),
with the same wave-numbers but different directions. As we shall see below,
these conclusions depend on the symmetry of the mixture and its
incompressibility.

\subsection*{Effects of asymmetry}
\label{sec:hpb-asymmetric-microphase-sep} 
The most important effect of the asymmetry is a third-order term in the Landau
expansion, rendering both the macrophase separation of the uncrosslinked
liquid and the microphase separation of the gel first-order transitions.

To keep the discussion simple, we neglect deviations of $k^2$ from $\kcrit^2$
and drop the $k$-dependence of the higher-order terms in the Landau free
energy. With the ansatz $\Psi^\alpha_{\vec{k}}=
(1-\delta_{\alpha,0})\,\Psi_{\vec{k}}$, corresponding to the mixed state in
the preparation ensemble and replica-symmetric phase separation in the
measurement ensemble, this leads to the free energy density
\begin{widetext} \begin{multline} \label{eq:282}
  \frac{f\left(\{\Psi\}\right)}{1-q^2} \ = \frac{1}{2} \plsum{\vec{k}}
  \lambda_2(k)\, \Psi_{\vec{k}}\Psi_{-\vec{k}} +\frac{q}{3}
  \plsum{\vec{k}_{1,2,3}} \Psi_{\vec{k}_1} \Psi_{\vec{k}_2} \Psi_{\vec{k}_3}
  \cdot\,\delta_{\vec{k}_1+ \vec{k}_2+ \vec{k}_3, \vec{0}} \\
  +\frac{q^2}{2} \bigg( \plsum{\vec{k}} \Psi_{\vec{k}} \Psi_{-\vec{k}}
  \bigg)^2 +\frac{1-3q^2}{12} \plsum{\vec{k}_{1,2,3,4}} \Psi_{\vec k_1}
  \Psi_{\vec{k}_2} \Psi_{\vec{k}_3} \Psi_{\vec{k}_4} \cdot
  \delta_{\vec{k}_1+\vec{k}_2+\vec{k}_3+\vec{k}_4,\vec{0}}\,,
\end{multline} \end{widetext}
with $\lambda_2(k)$ defined in eq.~(\ref{eq:42}).

Besides the lamellar microphases already discussed in
section~\ref{sec:mps-symmetric-case}, we now consider two additional
morphologies: Cylindrical phases having parallel orientation, aligned on a
honeycomb lattice in the perpendicular plane, and spherical domains on a body
centered cubic lattice. Although a randomly crosslinked blend will probably
reveal only local order, the regular structures are useful for constructing a
simple and tractable ansatz for the microphase separated state:
\begin{equation}
  \label{eq:283}
  \Psi_{\vec{p}} = \frac{\Psi}{\sqrt{m}}\, \sum_{i=1}^m\,
  (\delta_{\vec{p}\,,\,k\vec{n}_i}
  +\delta_{\vec{p}\,,\,-k\vec{n}_i}) 
\end{equation}
with $m=1$ for lamell\ae, $m=3$ for hexagonally ordered cylinders, and $m=6$
for spheres on a bcc lattice, and the corresponding lattice vectors
$\{\vec{n}_i\}$ being defined in
appendix~\ref{sec:rcp-mps-lattice-structures}.

With the lattice ansatz~(\ref{eq:283}), the evaluation of the higher-order
sums in eq.~(\ref{eq:282}) amounts to counting the number of possible \lqq
loops\rqq of two, three and four lattice vectors that add to zero. This is
carried out in appendix~\ref{sec:morphologies-wave-vectors-sums}, yielding
$\psum{\vec{k}} \Psi_{\vec{k}} \Psi_{-\vec{k}} = 2\Psi^2$, independent of the
morphology, and also
\begin{align}
  \label{eq:284}
  \plsum{\vec{k}_{1,2,3}} \Psi_{\vec{k}_1} \Psi_{\vec{k}_2} \Psi_{\vec{k}_3}
  \cdot\,\delta_{\vec{k}_1+ \vec{k}_2+ \vec{k}_3, \vec{0}} &= c_3^{(m)}
  \Psi^3,
  \\
  \label{eq:286}
  \plsum{\vec{k}_{1,2,3,4}} \Psi_{\vec k_1} \Psi_{\vec{k}_2} \Psi_{\vec{k}_3}
  \Psi_{\vec{k}_4} \cdot
  \delta_{\vec{k}_1+\vec{k}_2+\vec{k}_3+\vec{k}_4,\vec{0}} &= c_4^{(m)} \Psi^4,
\end{align}
where
\begin{align}
  \label{eq:287}
  c_3^{(1)}&=0, & c_3^{(3)}&=4/\sqrt{3}, & c_3^{(6)}&=4\cdot\sqrt{2/3}, \\
  c_4^{(1)}&=6, & c_4^{(3)}&=10, & c_4^{(6)}&=15.
\end{align}
Thus, the free-energy density becomes
\begin{multline}
  \label{eq:288}
  \frac{f(\{\Psi\})}{1-q^2}\,=\,
  \frac{\chimeas^{-1}-\chicrit^{-1}}{1-q^2}\, \Psi^2 +\frac{q
    c_3^{(m)}\!}{3}\, \Psi^3 \\
  +\left(2q^2 +\frac{(1-3q^2)c_4^{(m)}}{12}\right) \, \Psi^4 \,.
\end{multline}
Here, $\lambda_2$ has been evaluated at $\kcrit$ because (in this section) we
are not considering deviations of the wave-number from its critical value.

For $m=1$, the third-order term vanishes, even if $q\neq0$, so the transition
remains second order and the spinodal indeed indicates the equilibrium phase
transition point with respect to lamell\ae.  In contrast, for cylinders and
bcc spheres the equilibrium transition point~$\chitrans$ is shifted according
to
\begin{multline}
  \label{eq:290}
  \frac{1}{(1-q^2)\chitrans}-\frac{1}{(1-q^2)\chicrit} \\[1ex]
  =\frac{\big(qc_3^{(m)}/3\big)^2}{4\big(2q^2 +(1-3q^2)c_4^{(m)}/12\big)}
  \\[1ex] 
  =\frac{\big(qc_3^{(m)}\big)^2}{72q^2+3(1-3q^2)c_4^{(m)}} \\*[1ex]
  =\left\{ \begin{array}{l@{}ll} \displaystyle
      \frac{8q^2}{45-27q^2}&=\displaystyle
      \frac{24}{135}\,q^2+\mathcal{O}(q^4) & \ \genfrac{}{}{0pt}{0}{\text{for $m=3$}}{\text{(cylinders)}} \\[3.5ex]
%      \quad\text{for $m=3$ (cylinders),}\\[1.5ex]
      \displaystyle \frac{32q^2}{135-189q^2}
      &=\displaystyle\frac{32}{135}\,q^2+\mathcal{O}(q^4) &
      \ \genfrac{}{}{0pt}{0}{\text{for $m=6$}}{\text{(bcc spheres)}}  
%\quad\text{for $m=6$ (bcc spheres).}
  \end{array} \right.
\end{multline}
In the asymmetric case, the bcc spheres yield the lowest equilibrium
transition point of the three possibilities considered, \ie microphases first
occur with bcc symmetry.  This is to be expected, as the ratio of surface to
volume of the minority phases is minimal for spheres embedded in the majority
phase, and is in agreement with the finding of Alexander and
McTague~\cite{alex78} of a general preference for bcc symmetry in crystal
nucleation.
Note, however, that the Landau expansion is only valid for small $q^2$, for
which the transition is weakly first order. In particular, for $q^2=5/7$,
where the right hand side in the last line of eq.~(\ref{eq:290}) diverges, the
Landau expansion breaks down.

\subsection*{Effects of compressibility}
\label{sec:hpb-effects-compressibility} 

\begin{figure}
  \centering \psfrag{Y}{\large$\Psi(x)$} \psfrag{W}{\large$\rho(x)$}
  \hspace*{-1ex}
  \includegraphics[width=0.6\maxfigwidth]{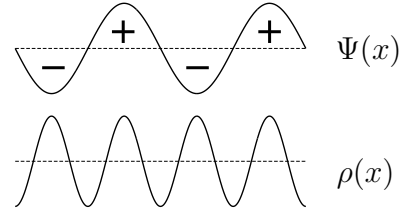}
  \caption{Coupling of mass ($\rho$) and charge ($\Psi$) density. Zones of
    large charge-density modulation are condensed, hence the mass-density is
    modulated with half the wave-length of the charge-density modulations.}
  \label{fig:hpb-coupling-psi-omega}
\end{figure}

A compressible system can avoid unfavorable {\itshape A-B\/} contacts and
lower its energy by diluting mixed regions having many such contacts, which
are characterized by a small absolute charge density, and condensing regions
that are rich in either {\itshape A} or {\itshape B}, which have a high
absolute charge density.  Mathematically, this becomes apparent via a nonzero
value of the saddle-point of the density $\bar{\rho}$ in a phase separated
state. For simplicity, we restrict the discussion of compressibility effects
to the symmetric case, where the shifted and the original density fields
coincide. In this case, the saddle point of $\rho$ is given by
\begin{equation}
  \label{eq:118}
  \bar{\rho}^\alpha_{\vec{k}} \,=\,
  \frac{\I}{2\big(1/\tilde{\lambda}^\alpha + g_\textup{D}(k^2)\big)}
  \psum{\vec{k_{1,2}}}  \Psi^\alpha_{\vec{k}_1}
  \Psi^\alpha_{\vec{k}_2} \cdot 
  \delta_{\vec{k}+\vec{k}_1+\vec{k}_2,\vec{0}}.
\end{equation}
For the simple example of lamellar microphases described by a single
wave-vector~$\vec{k}$ as in eq.~(\ref{eq:lamella}), eq.~(\ref{eq:118})
predicts density-field modulations having wave-vector $\vec{k}_1=\pm2\vec{k}$,
\ie with twice the wave-number of the charge-density modulations. This is
intuitively clear: Along one spatial period of the charge-density modulations,
their modulus or square, and thus the mass density, oscillates twice,
corresponding to the half wave-length or the double wave-number; this is
illustrated in fig.~\ref{fig:hpb-coupling-psi-omega}.

The compressibility is controlled by the strength of the excluded volume
interaction, \ie the parameter $\tilde{\lambda}=\lambda_m-\mu/V^n$.  To study
microphase separation in the symmetric but compressible case, we integrate out
the density field, keeping $\tilde{\lambda}$ finite.  With the ansatz
$\Psi^\alpha=(1-\delta_{\alpha,0}) \Psi$, we obtain
\begin{multline} 
  \label{eq:296}
  f\big(\Psi\big) = \frac{1}{2} \psum{\vec{k}}
  \lambda_2(k)\, \Psi_{\vec{k}} \Psi_{-\vec{k}} \\
  +\frac{1}{8\,\lambdaeff} \left(\psum{\vec{k}} \Psi_{\vec{k}}
    \Psi_{-\vec{k}} \right)^{\!2} \\*
  +\frac{1}{12} \left(1- \frac{3}{2\,\lambdaeff}\right)
  \psum{\vec{k}_{1,2,3,4}} \Psi_{\vec{k}_1} \Psi_{\vec{k}_2} \Psi_{\vec{k}_3}
  \Psi_{\vec{k}_4} \\ \times
  \delta_{\vec{k}_1+\vec{k}_2+\vec{k}_3+\vec{k}_4,\vec{0}}\,.
\end{multline}
Here, $\lambdaeff=\lambda-\mu+1$, and we have dropped the $k$ dependence in
the higher-order vertices, thereby restricting the domain size to its critical
value, determined by $\kcrit$ or the localization length of the gel.

\begin{figure}
  \hfill
  \includegraphics[width=0.23\columnwidth]{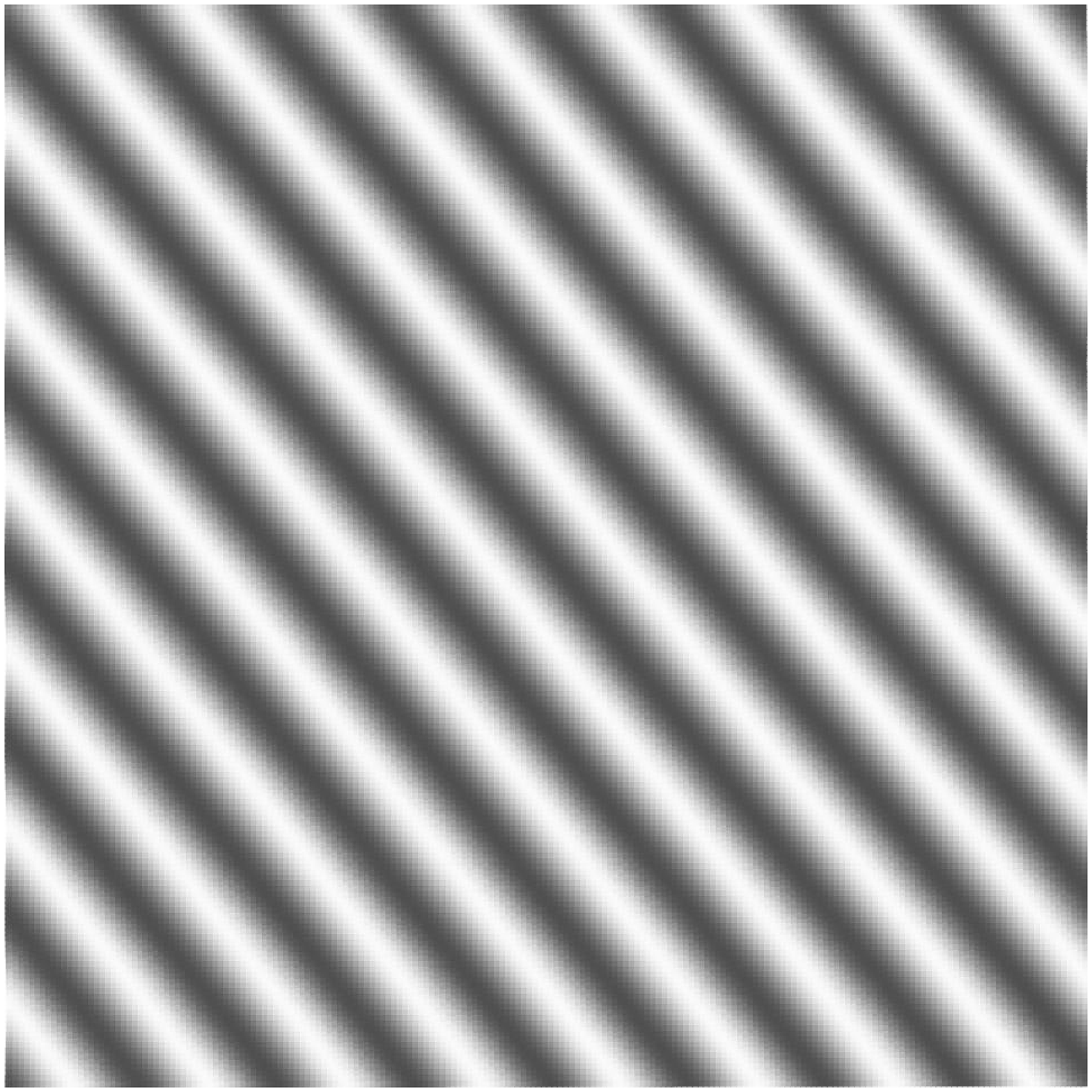}
  \hfill
  \includegraphics[width=0.23\columnwidth]{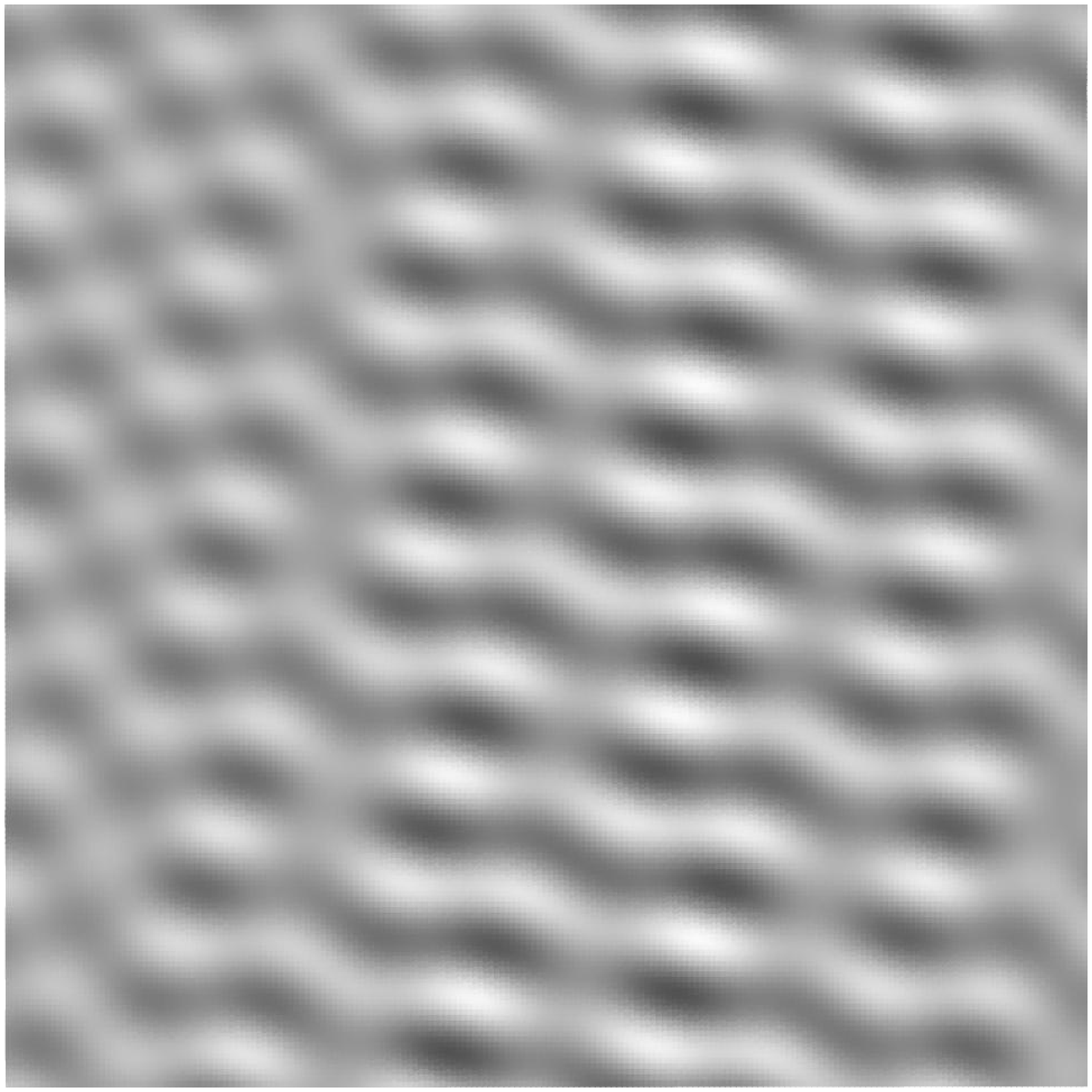}
  \hfill
  \includegraphics[width=0.23\columnwidth]{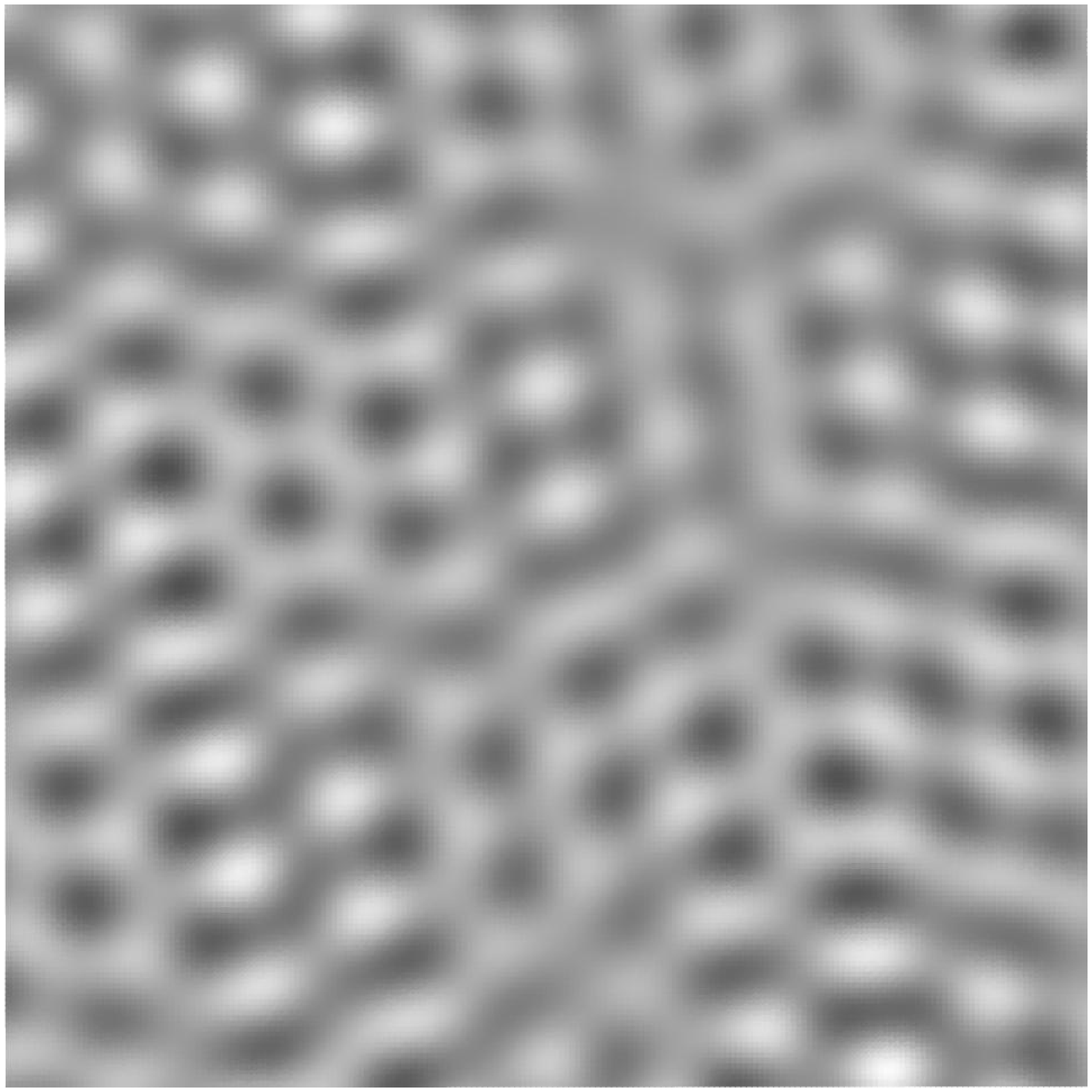}
  \hfill
  \includegraphics[width=0.23\columnwidth]{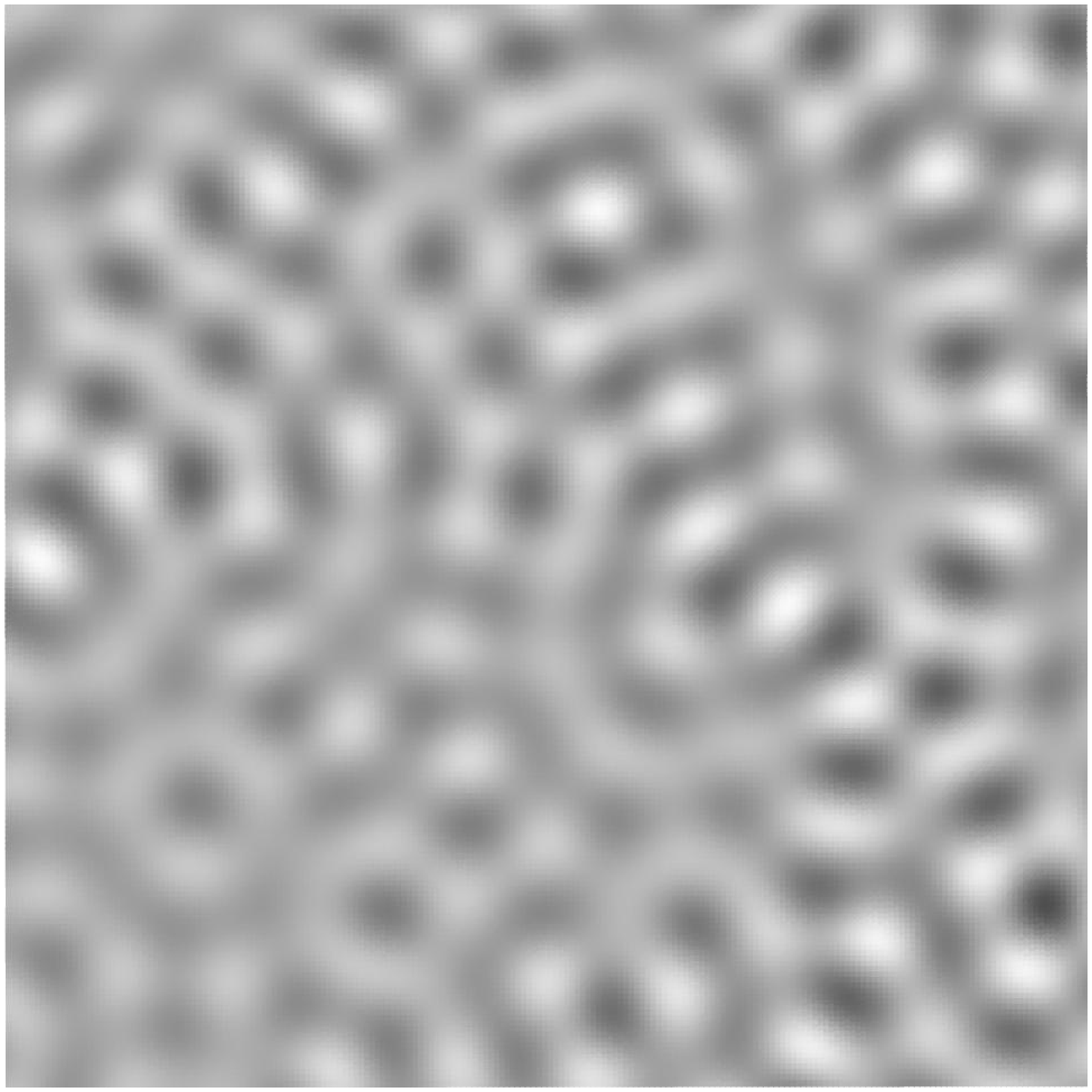} \hfill
  
  \caption{Superposition of lamell\ae\ having random orientations in real
    space in two dimensions. The pictures show an area of 10 by 10
    wave-lengths, and the local amplitude is indicated by the grey scale in
    arbitrary units for $Z=1,5,10$ and $100$ orientations.}
  \label{fig:hpb-random-morphologies}
\end{figure}

To account for the potential randomness of the microphase pattern, we extend
the previous lamellar ansatz by allowing a superposition of~$Z$
one-dimensional waves, each with identical wave-number~$\kcrit$ but random
phases~$\Phi_z$ and wave-vector orientations~$\vec{n}_z$, \ie,
\begin{equation}
  \label{eq:298}
  \Psi_{\vec{k}} = \frac{\Psi}{\sqrt{Z}} \sum_{z=1}^Z
  \left( \ee^{\I\,\Phi_z}\delta_{\vec{k},-\kcrit\vec{n}_z} 
    +\ee^{-\I\,\Phi_z}\delta_{\vec{k},\kcrit\vec{n}_z} \right),
\end{equation}
corresponding to $2\Psi/(VZ^{1/2})\cdot \sum_{z=1}^Z
\cos\left(\kcrit\vec{n}_z\vec{x} +\Phi_z\right)$ in real space. The optimal
number of orientations will be determined later. A few examples of such random
morphologies are shown in fig.~\ref{fig:hpb-random-morphologies}, the number
of phases ranging from 1 to 100.

We assume that none of the orientations $\vec{n}_z$ are collinear, and thus
the quadratic sums in eq.~(\ref{eq:296}) yield
\begin{align}
  \label{eq:299}
  \psum{\vec{k}} \lambda_2(\kcrit) 
  \Psi_{\vec{k}} \Psi_{-\vec{k}} &= 2\lambda_2(\kcrit)\,\Psi^2 %
  &\text{and} \\
  \psum{\vec{k}} \Psi_{\vec{k}} \Psi_{-\vec{k}} &= 2\Psi^2.
\end{align}
To compute the fourth-order sum in eq.~(\ref{eq:296}), we have to count the
number of possible closed loops of orientations. Because of the randomness,
the existence of quadruples of orientations able to form a closed loop is very
unlikely, except for the degenerate planar case of pairs of opposite vectors
$(\pm\vec{n}_{z}, \pm\vec{n}_{z^\prime})$, and hence we disregard non-planar
loops.  Single orientations allow for the construction of quadruples
$(\vec{n}_z,\vec{n}_z,-\vec{n}_z,-\vec{n}_z)$ that can be ordered in
$\binom{4}{2}=6$ ways. Quadruples $(\pm\vec{n}_z,\pm\vec{n}_{z^\prime})$ of
two pairs of different orientations can be ordered in $4!=24$ different ways,
and there are $\tfrac{1}{2}Z(Z-1)$ such pairs. Thus, the quartic sum in
eq.~(\ref{eq:296}) yields
\begin{multline}
  \label{eq:138}
  \psum{\vec{k}_{1,2,3,4}} \Psi_{\vec{k}_1} \Psi_{\vec{k}_2} \Psi_{\vec{k}_3}
  \Psi_{\vec{k}_4} \cdot \delta_{\vec{k}_1 +\vec{k}_2 +\vec{k}_3 +\vec{k}_4,
    \vec{0}} \\ =\frac{12Z(Z-1)+6Z}{Z^2}\, \Psi^4
  =12\left(1-\frac{1}{2Z}\right) \Psi^4.
\end{multline}

Inserting the sums into the free-energy density we obtain
\begin{multline}
  \label{eq:300}
  f(\Psi) 
  = \lambda_2(k)\,\Psi^2 \\ +\bigg(\!\Big(1-\frac{1}{\lambdaeff}\Big) -
  \Big(1-\frac{3}{2\lambdaeff}\Big)\frac{1}{2Z} \bigg)\Psi^4.
\end{multline}
The fourth-order term depends on $\lambdaeff$ and the number of components
$Z$. It has to be positive to guarantee stability, and this requires
$\lambdaeff>1$.  The sign of the $\mathcal{O}(Z^{-1})$ term determines the
optimal number of random orientations. For low compressibilities, \ie for
$\lambdaeff>3/2$, the term is negative and the free energy grows with an
increasing number of orientations, hence the simple lamellar morphology is
favored. For a rather compressible system having, in contrast,
$1<\lambdaeff<3/2$, the effective free energy decreases with increasing~$Z$,
favoring an \lqq infinite\rqq number of orientations and hence a random
pattern.

\section{Conclusions}

In this paper we have analyzed a microscopic model of crosslinked polymer
blends, built on the Edwards model for a polymer melt and generalized to two
components, which are mutually incompatible. Random crosslinks are introduced
according to the Deam-Edwards distribution, also generalized to include
concentration fluctuations at the instant of crosslinking. Thereby, the
concentration fluctuations in the melt are partially frozen in and sustained
in the gel phase. Apart from these correlations, which are present at
preparation, the crosslinks are taken to connect monomers irrespective of
their charge.  Hence, within mean-field theory, the resulting gel is identical
to the one made from just one species of polymer. However, concentration
fluctuations are present, and have been computed on the Gaussian level of
approximation. Of particular interest are the frozen-in or glassy
fluctuations, which reflect the preparation state.  In general, the network
can only quench fluctuations on length-scales larger than its own mesh size,
which is roughly given by the localization length of mean-field theory. If the
preparation ensemble is close to macroscopic phase separation then the
length-scale of these fluctuations is large compared to the mesh size, so that
the glassy fluctuations are given by the concentration fluctuations in the
preparation state. If, on the other hand, the preparation state is far from
phase separation, the frozen-in charge fluctuations follow the network
pattern, and hence are completely random, because the crosslinks are not
sensitive to the species. The thermal concentration fluctuations are
independent of the preparation state.

Lowering the temperature in the gel, or equivalently increasing the
incompatibility of the two species, gives rise to phase separation, the
spatial extent of which is limited by the mesh size of the network.  The
length-scale of the resulting ``micro''-phases can thus range from almost
microscopic to nearly macroscopic scales, depending on the degree of
crosslinking of the gel.  The instability towards microphase separation is
signaled by a divergence of the time-persistent as well as the thermal
fluctuations. The emergent microstructure is shown to depend sensitively on
charge imbalance and compressibility. The latter allows for random patterns
with a unique wave-length, \ie, the localization length, whereas in the
incompressible system lamell\ae\ are favorable for balanced mixtures and
hexagonal patterns for imbalanced mixtures.

We now compare our results to previous phenomenological approaches, many of
which have focused on the issue of microphase separation. de Gennes argued
that the charges in the crosslinked gel cannot move freely but are displaced
in analogy to the charges in a dielectric material. He introduced a
polarization ${\vec P}$ which, as in electrostatics, is determined by the
charges according to $\nabla \cdot {\vec P}=-\Psi$. In the limit of weak
segregation, the free energy is quadratic in the polarization, and is simply
added to the free energy of charge fluctuations, resulting in
\begin{equation}
  f\left(\{\Psi\}\right)  = \frac{1}{2} \plsum{\vec{k}} 
  \left(\chicrit-\chimeas+k^2+\frac{C}{k^2}\right)\Psi_{\vec{k}}
  \Psi_{-\vec{k}}\,,
\end{equation}
where $C$ is a coefficient of ``internal rigidity''. The above free energy
leads to an instability at finite wave-number (microphase separation), but
predicts that $\lim_{k\to 0} \Ssc(k)=0$, in disagreement with experiment. The
nonzero scattering intensity at zero wave-number is due to the frozen-in
charge fluctuations present at preparation. To account for these fluctuations,
Benhamou \etal~\cite{benm94} refined the analogy to a dielectric by including
a Debye-H\"uckel screening of the ``charges'', which permits long-range
inhomogeneities leading to a non-vanishing zero-angle scattering. The
screening length $\kappa$ is determined self-consistently, by assuming that the
scattering intensity at $k=0$ is not affected by the crosslinking as long as
the temperature remains unchanged after preparation~\cite{bett95}. The free
energy in the quadratic approximation then reads:
\begin{equation}
  f\left(\{\Psi\}\right)  = \frac{1}{2} \plsum{\vec{k}} 
  \left(\chicrit-\chimeas+k^2+\frac{C}{k^2+\kappa^2}\right) \Psi_{\vec{k}}
  \Psi_{-\vec{k}} \,.
\end{equation}
This expression can be compared with eq.~(\ref{eq:282}) in the quadratic
approximation:
\begin{widetext} \begin{align}
    f\left(\{\Psi\}\right) &= \frac{1}{2} \plsum{\vec{k}}
    \left(\frac{1}{\chimeas}-g_\textup{D}(k^2)+\mu(\mu-1)\omega(k^2(\mu-1))
    \right)\, \Psi_{\vec{k}}\Psi_{-\vec{k}} \\
    &\approx \frac{1}{2} \plsum{\vec{k}}
    \left(1-\chimeas+k^2\frac{\chimeas}{3}+\mu(\mu-1)\omega(k^2(\mu-1))
    \right)\, \Psi_{\vec{k}}\Psi_{-\vec{k}}\,.
\end{align} \end{widetext}
In the last line, we have expanded the Debye function for small wave-number.
We see that the microscopic model indeed agrees with phenomenological
theories, provided we identify the phenomenological terms with the order
parameter of the gel. The wave-number dependence of the order parameter is not
a Lorentzian; nevertheless, it decays monotonically with $k$, the relevant
length-scale being given by the localization length. Hence, the somewhat
mysterious screening length is unambiguously identified with the localization
length, which is computed self-consistently. Thereby, the microscopic model
substantiates the picture of de Gennes and, furthermore, allows the
computation of the parameters and functions that are beyond the
phenomenological approach.

The frozen-in fluctuations were first addressed by Read \etal~\cite{read95},
who considered a blend of polymer chains anchored at both ends to randomly
chosen fixed points in space, in order to account approximately for the
localization of chains due to the crosslinks.  Read~\etal make reasonable but
\emph{ad hoc} assumptions about the distribution of the quenched random
end-to-end vectors, and solve the resulting model within the random phase
approximation. They obtain a scattering function that exhibits a nonzero value
at $k=0$, due to the random, quenched fluctuations. In addition they compute
the thermal, as well as the glassy, charge fluctuations:
\begin{align}
  \Svr(k)&=\frac{1}{\chicrit-\chi+k^2+\frac{C}{k^2}}& \text{and}\\
  \Sgl(k)&=\frac{(\frac{C}{k^2})^2|\rho_0(k)|^2}
  {(\chicrit-\chi+k^2+\frac{C}{k^2})^2}\,;
\end{align}
where $\rho_0(k)$ is the frozen-in concentration.  The above results are in
close correspondence to the results of our analysis, presented in
eqs.~(\ref{eq:33},\ref{eq:37}) and evaluated in the limit of small wave-number
\begin{equation*}
  \Svr(k) \sim \frac{1} {1-\chimeas+k^2\frac{\chimeas}{3}+\chimeas
  (\mu-1)\omega(k^2/(\mu-1))}\,,
\end{equation*}
\begin{equation*}
  \Sgl(k) \sim \frac{(\mu-1)\scalf(k^2/(\mu-1)}
  {(1-\chimeas+k^2\frac{\chimeas}{3}+\chimeas
    (\mu-1)\scalf(k^2/(\mu-1)))^2}\,.
\end{equation*}
Both approaches, the phenomenological one and the microscopic model, predict a
divergence as microphase separation is approached, with the glassy correlations
diverging twice as strongly as the thermal ones.

The work reported in the present paper can be extended in several directions.
First, we have worked only on the level of mean-field and Gaussian
fluctuations.  It is known that the microphase separation transition in the
symmetric case is rendered first order by fluctuations~\cite{braz75}, and
hence it would be interesting to see the effect of fluctuations, even though
the critical region is expected to be small~\cite{benh96}.  Another extension
is a crosslink probability that depends on the species.  This would allow us
to study, among other things, interpenetrating networks. Finally, it would be
interesting to look at the dynamics of microphase separation.

%\clearpage
\appendix

\section{Microphase morphology}
\label{sec:rcp-mps-morphology}
To investigate microphase transition, we assume that the phase-separation
pattern can be described by a first-harmonic ansatz having a dominant
wave-number $k$ and a definite lattice structure:
\begin{equation} \label{eq:270}
  \Psi_{\vec{k}^\prime} = \Psi\,\frac{V}{\sqrt{2m}}\,
  \sum_{i=1}^m \left( \delta_{\vec{k}^\prime,+k\vec{n}_i}
  +\delta_{\vec{k}^\prime,-k\vec{n}_i} \right)\,, 
\end{equation}
with lattice vectors $n_i\in\mathbb{G}:=\{\vec n_i\,|\,i=1\dots m\}$ and an
amplitude $\Psi$. 

\subsection{Lattice structures} \label{sec:rcp-mps-lattice-structures}
We consider three particular morphologies, which are known to occur in the
microphase separation of regular copolymer melts~\cite{leib80,maye89}:
\begin{description}
\item[lamell\ae\ $(m=1)$:] Alternating sheets rich in $A$ and $B$;
  one-dimensional order. Lattice vector: $\vec n_1=(1,0,0)^T$.
\item[cylinders $(m=3)$:] Close-packed, \ie~hexagonally arranged, cylindrical
  domains, $A$ in $B$ or vice versa; two-dimensional order. Lattice vectors:
  $\vec{n}_1=(1,0,0)^T$, $\vec{n}_2=(-1/2,\sqrt3/2,0)^T$ and
  $\vec{n}_3=(-1/2,-\sqrt3/2,0)^T$.
\item[bcc spheres $(m=6)$:] Spherical $A$-rich domains in $B$, or vice versa,
  on a bcc lattice in real space; three-dimensional order.
  Lattice vectors of the corresponding fcc lattice in Fourier space:
  $\vec{n}_1=(1,1,0)^T/\sqrt{2}$, $\vec{n}_2=(0,1,1)^T/\sqrt{2}$,
  $\vec{n}_3=(1,0,1)^T/\sqrt{2}$, $\vec{n}_4=(1,0,-1)^T/\sqrt{2}$
  $\vec{n}_5=(-1,1,0)^T/\sqrt{2}$ and $\vec{n}_6=(0,-1,1)^T/\sqrt{2}$.
\end{description}
Note that $\vec a\in\mathbb{G}\Rightarrow-\vec a\notin\mathbb{G}$; therefore
we introduce the symmetrized set of lattice vectors, $\mathbb{G}^+:=\{\vec
n\,|\,\vec n\in \mathbb{G}\vee-\vec n\in\mathbb{G}\}$.  For $m>1$, the set
$\mathbb{G}$ of lattice vectors is not minimal in the sense of linear
independence: for any two vectors $\vec a\neq\vec b\in\mathbb{G}$, the
difference $\vec a-\vec b$ is included in $\mathbb{G}^+$. Rather, the vectors
are chosen such that the $k_i\in\mathbb{G}^+$ point towards the directions of
all nearest-neighbor lattice sites.

\subsubsection{Wave-vector sums} \label{sec:morphologies-wave-vectors-sums} 
Inserting the ansatz (\ref{eq:270}) into the expansion of the Landau free
energy of the random copolymer melt or the crosslinked homopolymer blend
yields sums of the type
 \begin{widetext} \begin{multline}
   \label{eq:273}
   \psum{\vec{k}_{1,\ldots,p}} f_2(\vec{k}_1,\ldots,\vec{k}_{p-1})
   \Psi_{\vec{k}_1} \cdots \Psi_{\vec{k}_p} \delta_{\sum_{\nu=1}^p
     \vec{k}_\nu,\vec{0}} \\=\frac{\Psi^p\cdot V^p}{(2m)^{p/2}}
   \psum{\vec{k}_{1,\ldots,p}} f_2(\vec{k}_1,\ldots,\vec{k}_{p-1})
   \prod_{\nu=1}^p \left(\sum_{\mu=1}^m (\delta_{\vec{k}_\nu,+k\vec{n}_\mu}
     +\delta_{\vec{k}_\nu,-k\vec{n}_\mu})\!
   \right) \delta_{\sum_{\nu^\prime=1}^p \vec{k}_{\nu^\prime},\vec{0}} \\
   =\Psi^p\cdot \frac{V^p}{(2m)^{p/2}} \sum_{\vec{k}_{1,\ldots,p}
     \in\mathbb{G}^+} f_2(\vec{k}_1,\ldots,\vec{k}_{p-1})\,
   \delta_{\vec{k}_1+\ldots+\vec{k}_p,\vec{0}}\,.
 \end{multline} \end{widetext}
%
%% \begin{multline}
%%   \label{eq:273}
%%   \psum{\vec{k}_{1,\ldots,p}} f_2(\vec{k}_1,\ldots,\vec{k}_{p-1})
%%   \Psi_{\vec{k}_1} \cdots \Psi_{\vec{k}_p} \delta_{\sum_{\nu=1}^p
%%     \vec{k}_\nu,\vec{0}} \\[1ex]
%%   \shoveleft{=\frac{\Psi^p\cdot V^p}{(2m)^{p/2}} \psum{\vec{k}_{1,\ldots,p}}
%%     f_2(\vec{k}_1,\ldots,\vec{k}_{p-1})} \\ \shoveright{\times \prod_{\nu=1}^p
%%     \left(\sum_{\mu=1}^m (\delta_{\vec{k}_\nu,+k\vec{n}_\mu}
%%       +\delta_{\vec{k}_\nu,-k\vec{n}_\mu})\!
%%     \right) \delta_{\sum_{\nu^\prime=1}^p \vec{k}_{\nu^\prime},\vec{0}}} \\
%%   =\frac{\Psi^p\cdot V^p}{(2m)^{p/2}} \sum_{\vec{k}_{1,\ldots,p}
%%     \in\mathbb{G}^+}\!\!\!\! f_2(\vec{k}_1,\ldots,\vec{k}_{p-1})\,
%% \delta_{\sum_{\nu^\prime=1}^p \vec{k}_{\nu^\prime},\vec{0}}
%% %  \delta_{\vec{k}_1+\ldots+\vec{k}_p,\vec{0}}\,.
%% \end{multline}

In the quadratic sum the vertex function can be factored out, so that
\begin{multline}
  \label{eq:274}
  \frac{1}{V^2} \sum_{\vec{k}^\prime} f({k^\prime}^2)\Psi_{\vec{k}^\prime}
  \Psi_{-\vec{k}^\prime} \\
  =\Psi^2\cdot \frac{f(k^2)}{2m}\, \sum_{\vec{k_{1,2}} \in\mathbb{G}^+}
  \delta_{\vec{k}_1+\vec{k}_2,\vec{0}} = \Psi^2\cdot f(k^2)\,.
\end{multline}
The higher-order sums, however, strongly depend on the morphology of the
microphases. In the simplest case, $f(\vec{k}_1,\ldots)=1$, the computation of
these loops amounts to counting the number of closed loops that can be
constructed with the vectors in $\mathbb{G}^+$. In general, the loops must
also be classified with respect to their shape, \ie planar or non-planar, as
distinct shapes yield distinct values of the vertex functions. The counting
and classification have been done, \eg, in~\cite{maye89}, with the results
shown in table~\ref{tab:loop-counting}.
\begin{table}
  \begin{center}
    \begin{tabular}{llrrr}
       \multicolumn{2}{l}{loop type} & \multicolumn{1}{c}{lamell\ae} &
      \multicolumn{1}{c}{cylinders} & \multicolumn{1}{c}{bcc spheres} 
      \\[0.2ex] \hline 2-loop && 2$\quad$ & 6$\quad$ & 12$\quad$ \\[0.7ex]
      3-loop && 0$\quad$ & 12$\quad$ & 48$\quad$ \\[0.7ex]
      4-loop & planar & 6$\quad$ & 90$\quad$ & 396$\quad$ \\
      & nonplanar & --$\quad$ & --$\quad\,$ & 144$\quad$ \\
      & total & 6$\quad$ & 90$\quad$ & 540$\quad$
    \end{tabular}
    \label{tab:loop-counting}
    \caption{Number of closed loops of $p$ lattice vectors (\lqq$p$-loops\rqq)
      for different morphologies. The $4$-loops are divided into intra- and
      extra-planar loops.}
  \end{center}
\end{table}
For $f(\vec{k},\ldots)=1$, the third- and fourth-order sums are given by
\begin{multline} 
  \label{eq:275}
  \frac{1}{V^3} \sum_{\vec{k}_{1,2}} \Psi_{\vec{k}_1} \Psi_{\vec{k}_2}
  \Psi_{\!-\vec{k}_1-\vec{k}_2}\\
  =\Psi^3\cdot \begin{cases}
    0&\text{for $m=1$ (lamell\ae),} \\
    \sqrt{2/3}&\text{for $m=3$ (cylinders),} \\
    2/\sqrt{3}&\text{for $m=6$ (bcc spheres),}
  \end{cases}
\end{multline}
and
\begin{multline} 
  \label{eq:276}
  \frac{1}{V^4} \sum_{\vec{k}_{1,2,3}} \Psi_{\vec{k}_1} \Psi_{\vec{k}_2}
  \Psi_{\vec{k}_3} \Psi_{\!-\vec{k}_1-\vec{k}_2-\vec{k}_3} \\
  =\Psi^4\cdot
  \begin{cases}
    3/2&\text{for $m=1$ (lamell\ae),} \\
    5/2&\text{for $m=3$ (cylinders),} \\
    15/4&\text{for $m=6$ (bcc spheres).}
  \end{cases}
\end{multline}

\section{Vertex functions} \label{sec:hpb-vertex-funct}
In the following, we define and compute the vertex functions appearing in the
Landau expansion of the effective free energy, which are integrals over Wiener
correlators of the type
\begin{multline}
  \label{eq:248}
  \bigg\langle \exp\bigg\{ -\I \sum_{\zeta=1}^z \hat{k}_\zeta
  \cdot\hat{r}(s_\zeta) \bigg\} \bigg\rangle^\textup{\!\!W}_{\!n} \\
  = \delta_{\sum_{\zeta=1}^z \hat{k}_\zeta,\hat{0}}\, \exp\bigg\{ \frac{1}{2}
  \sum_{\zeta,\zeta^\prime} |s_\zeta-s_{\zeta^\prime}|\, \hat{k}_\zeta \cdot
  \hat{k}_{\zeta^\prime} \bigg\};
\end{multline}
a derivation of eq.~(\ref{eq:248}) can be found in~\cite{gold96}. The
correlator vanishes unless the wave-vectors sum to zero in each replica. If
just single-replica quantities are involved, the correlator factorizes,
\begin{multline*}
  %\label{eq:247}
  \bigg\langle\! \exp\bigg\{ -\I \sum_{\zeta_1=1}^{z_1}
  \vec{k}_{\zeta_1}\cdot\vec{r}^{\alpha_1}(s_{\zeta_1}) -\ldots\\
  \shoveright{-\I \sum_{\zeta_m=1}^{z_m}
    \vec{k}_{\zeta_m}\cdot\vec{r}^{\alpha_m}(s_{\zeta_m})
    \bigg\} \bigg\rangle^\textup{\!\!W}_{\!n}} \\
  \shoveleft{= \bigg\langle\! \exp\bigg\{ -\I \sum_{\zeta_1=1}^{z_1}
    \vec{k}_{\zeta_1}\vec{r}(s_{\zeta_1}) \bigg\} \bigg\rangle^\textup{\!\!W}
    \times
    \cdots}\\ \times \bigg\langle\! \exp\bigg\{ -\I \sum_{\zeta_m=1}^{z_m}
    \vec{k}_{\zeta_m}\vec{r}(s_{\zeta_m}) \bigg\} \bigg\rangle^\textup{\!\!W}
\end{multline*} 
for pairwise distinct~$\alpha_1,\ldots,\alpha_m$, where
$\langle\,\cdots\,\rangle^\textup{\!W}$ denotes the unreplicated Wiener
average.

\subsection{Definition of the vertex functions}
\label{sec:corr-funct} 
The second-order coefficients of the Landau expansion are governed by the
\emph{Debye function}
\begin{multline*} 
%  \label{eq:158}
  g_\textup{D}\bigl( k^2 \bigr) := \int_0^1\!\D s_1\D s_2\,\Bigl\langle
  \ee^{-\I\vec{k}(\vec{r}(s_1)-\vec{r}(s_2))} \Bigr\rangle^\textup{\!\!W}
  \\
  =\int_0^1\!\D s_1\D s_2\,\Bigl\langle
  \ee^{-\I\hat{k}(\hat{r}(s_1)-\hat{r}(s_2))}
  \Bigr\rangle^\textup{\!\!W}_{\!n}\\
  =\frac{\ee^{-k^2}-1+k^2}{ k^4/2} % \\
  = 1-\tfrac{1}{3} k^2 +\tfrac{1}{12} k^4\ +\mathcal{O}(k^6),
\end{multline*}
the scattering function for a non-interacting Gaussian chain.

The third-order correlators,
\begin{multline*}
  g_3\big(\vec{k}_1, \vec{k}_2 \big) \\
  := \int_0^1\!\D s_1\D s_2\D s_3\, \left.  \Bigl\langle
    \ee^{-\I\sum_{\nu=1}^3 \vec{k}_\nu\vec{r}(s_\nu)}
    \Bigr\rangle^\textup{\!\!W} \right|_{\vec{k}_3=-\vec{k}_1-\vec{k}_2} \\
  = \int_0^1\!\D s_1\D s_2\D s_3\, \left. \Bigl\langle \ee^{-\I\sum_{\nu=1}^3
      \hat{k}_\nu\hat{r}(s_\nu)} \Bigr\rangle^\textup{\!\!W}
    \,\right|_{\hat{k}_3=-\hat{k}_1-\hat{k}_2}
\end{multline*}
and
\begin{widetext} 
\begin{multline*} %\label{eq:159}
  g_{\Psi^2\Omega}\bigl( \vec{k}_1, \vec{k}_2 \bigr) := \int_0^1\!\D s_1\D
  s_2\D s_3\, \Bigl\langle \ee^{-\I\vec{k}_1(\vec{r}(s_1)-\vec{r}(s_3))}
  \Bigr\rangle^\textup{\!\!W} \Bigl\langle
  \ee^{-\I\vec{k}_2(\vec{r}(s_2)-\vec{r}(s_3))}
  \Bigr\rangle^\textup{\!\!W}\\
  =\int_0^1\!\D s_1\D s_2\D s_3\, \left. \Bigl\langle
    \ee^{-\I\vec{k}_1\vec{r}^{\alpha_1}(s_1)
      -\I\vec{k}_2\vec{r}^{\alpha_2}(s_2) -\I\hat{k}\hat{r}(s_3)}
    \Bigr\rangle^\textup{\!\!W}_{\!n}
  \right|_{\dsubscr{\alpha_1\neq\alpha_2}{\hat{k}=-\vec{k}_1\otimes
      e_{\alpha_1}-\vec{k}_2\otimes e_{\alpha_2}}} \,,
\end{multline*} \end{widetext}
describe the correlation between three one-replica fields and the correlation
between two one-replica-fields and a higher-order replica field, respectively.

Finally, the fourth-order correlator is given by
\begin{multline*}
  g_{\Psi^4}\big(\vec{k}_1, \vec{k}_2, \vec{k}_3 \big) = \\
  \int_0^1\!\!\D s_1\D s_2\D s_3\D s_4\!\left. \Bigl\langle
    \ee^{-\I\sum_{\nu=1}^4 \vec{k}_\nu\vec{r}(s_\nu)}
    \Bigr\rangle^\textup{\!\!W} \right|_{\vec{k}_4=-\sum_{\nu=1}^3
    \vec{k}_\nu}.
\end{multline*}

\subsubsection{Lamellar case}

The third- and fourth-order correlators depend on the directions of the
wave-vectors. In particular, we require the \lqq lamellar\rqq case, in which
all wave-vectors are collinear. We note that $g_3(\vec{k},-\vec{k})=
g_\textup{D}(k^2)$, and define
\begin{widetext} \begin{align*}
    g_3(k^2) &:= g_3(\vec{k},\vec{k}) =\frac{-(\ee^{-4k^2}-1+4k^2-8k^4)
      +64(\ee^{-k^2}-1+k^2-\tfrac{1}{2}k^4)}{12k^6} +\frac{2(\ee^{-k^2}-1+k^2)}{k^4} \\
    &=1-k^2+\tfrac{3}{4}k^4+\mathcal{O}(k^6)\,, \\[1.5ex]
    g_{\Psi^2\Omega}(k^2) &:= g_{\Psi^2\Omega}(\vec{k}, -\vec{k}) =
    \frac{-\ee^{-2 k^2} +(8+2 k^2)\ee^{- k^2} +4 k^2-7}{k^6}
    =1-\tfrac23 k^2+\tfrac{17}{60} k^4 \ +\mathcal{O}(k^6) \\
    \intertext{and} g_{\Psi^4}\big( k^2 \bigr) &:=g_{\Psi^4}\big(\vec{k},
    \vec{k}, -\vec{k}\, \big) = \frac{144 k^4 -60 k^2(9+4\ee^{- k^2})
      +784(1-\ee^{- k^2})
      -(1-\ee^{-4 k^2})}{18 k^8} \\
    &= 1-\tfrac{2}{3}k^2\ +\tfrac{11}{30}k^4 +\mathcal{O}(k^4)\,.
\end{align*} \end{widetext}

\section{Scaling function for the gelation order parameter}
\label{sec:scaling-function} 
 
The localization lengths~$\tau$ of monomers in the gel fraction of a
crosslinked homopolymer melt or blend are distributed according to the
distribution~$p(\tau)$; see section~\ref{sec:crossl-homog-mixed}.  The
fraction of the gel and the distribution are determined from the
self-consistent solution of the saddle-point equations with the order
parameter hypothesis~(\ref{eq:21}). The gelation order parameter, essentially
the Laplace transform of~$p(\tau)$, is proportional to a scaling
function~$\omega(x)$, which is computed in~\cite{gold96} in the asymptotic
regimes of small and large~$x$. For convenience, we define a rescaled version
of~$\omega$,
\begin{equation}
  \label{eq:161}
  w(k^2) := 2\cdot\omega\big(\sqrt{8k^2/3}\,\big),
\end{equation}
This also absorbs a factor of two arising from the different length-scale used
in~\cite{gold96} (the Wiener Hamiltonian used therein differs by a factor of
two).

\begin{figure}
  \psfrag{PFo3}[Bc]{\large$\ \omega_\textup{ip}(x)$}
  \psfrag{PFo1}[Bc]{\large$\ \omega_1(x)$} \psfrag{PFo2}[Bc]{\large$\ 
    \omega_2(x)$} 
  
  \psfrag{PY0.0}[Bl]{\normalsize\ \ {$0.0$}} %
  \psfrag{PY0.2}[Bl]{\normalsize\ \ {$0.2$}}
  \psfrag{PY0.4}[Bl]{\normalsize\ \ {$0.4$}} %
  \psfrag{PY0.6}[Bl]{\normalsize\ \ {$0.6$}} %
  \psfrag{PY0.8}[Bl]{\normalsize\ \ {$0.8$}}
  \psfrag{PY1.0}[Bl]{\normalsize\ \ {$1.0$}} %
  \psfrag{PX0}[Bl]{\normalsize\ \,$0$} \psfrag{PX1}[Bl]{\normalsize\ \,$1$}
  \psfrag{PX2}[Bl]{\normalsize\ \,$2$} \psfrag{PX3}[Bl]{\normalsize\ \,$3$}
  \psfrag{PX4}[Bl]{\normalsize\ \,$4$} \psfrag{PFk}[cl]{\large\ $x$}

  \center
    \includegraphics[width=\maxfigwidth]{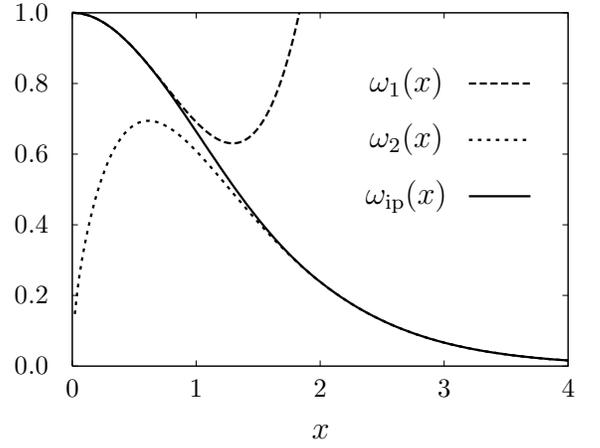}
  \caption[Scaling function]{
    Scaling function $\omega(k)$ versus~$k$: Asymptotic expressions and
    interpolated function.}
  \label{fig:scaling-function}
\end{figure}

\subsection{Interpolation formula} \label{sec:interp-form}
The scaling function $\omega(k)$ defined in~\cite{gold96} can be described
asymptotically by
\begin{equation*}
  \omega(x) \approx \begin{cases} \,
    \omega_1(x) := 1 - 0.4409x^2 + 0.1316x^4 & \text{for } x\ll1, \\[1ex] \,
    \omega_2(x) :=
    3\bigl(\frac{\pi^2 x^{6}}{8\cdot\,1.678}\bigr)^{\!1/4}\,
    \ee^{-\sqrt{2\cdot\,1.678}\,x} \\[.5ex]
    \hfill \times \Bigl(1+\frac{27}{40\sqrt{2\cdot1.678}\,x\,}\Bigr)
    & \text{for } x\gg1. \\
  \end{cases}
\end{equation*}
In order to access the whole range of $0<x<\infty$ we interpolate between the
asymptotic regimes using the interpolation formula
\begin{equation}
  \omega(k) \approx \begin{cases}
    \omega_1(x), & \text{for $x<\tfrac{1}{2}$}, \\
    \omega_\textup{ip}(x),
    & \text{for $\tfrac{1}{2}\leq x<2$,} \\
    \omega_2(x), & \text{for $x\geq2$,}
  \end{cases}
\end{equation}
with the interpolating rational function
\begin{equation}
  \label{eq:250}
  \omega_\textup{ip} := \frac{b_0+b_1x}{1+a_1x+a_2x^2+a_3x^3}\,.
\end{equation}
The coefficients $a_1=-0.055$, $a_2=0.165$, $a_3=0.139$, $b_0=1.023$ and
$b_1=-0.194$ are chosen such that the value and first derivative of
$\omega_\textup{ip}(x)$ coincide with those of $\omega_1(x)$ at $x=1/2$ and
with those of $\omega_2(x)$ at $x=2$; an additional sampling point is the
numerical value $\omega(x=1)=0.664$.

%-------------------------------------------------------------------------

\begin{acknowledgements}
  
  We thank Xiangjun Xing for enlightening discussions.  This work was
  supported in part by the Deutsche Forschungsgemeinschaft through SFB~602
  (AZ, CW), Grant No.~Zi~209/6-1 (AZ), and GRK~782 (CW), and by the
  U.S.~National Science Foundation through grant NSF DMR02-05858, and the U.S.
  Department of Energy through grant DEFG02-96ER45439 (PMG).

\end{acknowledgements}

%-------------------------------------------------------------------------

\newcommand{\dash}{-}
\newcommand{\jensuwe}{J.-U.\ }
\bibliography{HPB-condmat.bib}

\end{document}